\documentstyle[11pt,aaspp4,psfig]{article}   

\def\Lmin{$L_{\rm min}$}
\def\L1{$L_1$}
\def\L2{$L_2$}
\def\fL{$\phi (L)$}
\def\SL{${\cal S}(L)$}
\def\NL{$N(L)$}
\def\Ng{$N_G$}

\lefthead{Kalogera et al.\ }
\righthead{NS--NS Coalescence Rate}

\begin{document}


\title{The Coalescence Rate of Double Neutron Star Systems} 
\author{V.\ Kalogera\altaffilmark{1}, R.\ Narayan\altaffilmark{1}, 
D.N.\ Spergel\altaffilmark{2,3}, and J.H.\ Taylor\altaffilmark{4}}
\altaffiltext{1}{Harvard-Smithsonian Center for Astrophysics, 60
Garden St., Cambridge, MA 02138; vkalogera and rnarayan@cfa.harvard.edu}
\altaffiltext{2}{Princeton University Observatory, Princeton, NJ 08544;
dns@astro.princeton.edu}
\altaffiltext{3}{WM Keck Distinguished Visiting Professor
of Astrophysics, IAS, Princeton NJ 08540}
\altaffiltext{4}{Joseph Henry Laboratories and Physics Department,
Princeton University, Princeton, NJ 085544; joe@pulsar.princeton.edu}

\begin{abstract}

We estimate the coalescence rate of close binaries with two neutron stars
(NS) and discuss the prospects for the detection of NS--NS inspiral events
by ground--based gravitational--wave observatories, such as LIGO. We
derive the Galactic coalescence rate using the observed sample of close
NS--NS binaries (PSR~B1913+16 and PSR~B1534+12) and examine in detail each
of the sources of uncertainty associated with the estimate. Specifically,
we investigate (i) the dynamical evolution of NS--NS binaries in the
Galactic potential and the vertical scale height of the population, (ii)
the pulsar lifetimes, (iii) the effects of the faint end of the radio
pulsar luminosity function and their dependence on the small number of
observed objects, (iv) the beaming fraction, and (v) the extrapolation of
the Galactic rate to extragalactic distances expected to be reachable by
LIGO. We find that the dominant source of uncertainty is the correction
factor (up to $\simeq 200$) for faint (undetectable) pulsars. All other
sources are much less important, each with uncertainty factors smaller
than 2. Despite the relatively large uncertainty, the derived coalescence
rate is consistent with previously derived upper limits, and is more
accurate than rates obtained from population studies. We obtain a most
conservative conclusion that the detection rate for LIGO II lies in from
range from 2 events per year up to at least 300 events per year or even
possibly in excess of 1000 events per year.

\end{abstract}

\section{INTRODUCTION}

The importance of close binaries with two neutron stars (NS--NS) for
gravitational wave physics was established a few years after the discovery
of the prototype NS--NS system, the binary PSR~B1913+16 (\cite{HT75}1975),
with the measurement of orbital decay at a rate consistent with
gravitational wave emission as predicted by general relativity (at a
$3\times 10^{-3}$ accuracy level) (\cite{TW82}1982; 1989; 1999,
unpublished). This orbital decay is expected to end catastrophically with
the merger of the two neutron stars as the binary orbital separation
becomes comparable to the NS radii (for a recent review, see
\cite{RS99}1999). Such inspiral events and the final mergers could
possibly be detected as gravitational wave sources by the currently built
ground--based laser interferometers, such as LIGO and VIRGO. With the
upcoming completion of these observatories increased interest has focused
on close binaries with two compact objects.

Assessment of the detectability of binary compact inspiral events depends
both on the strength of the gravitational wave signal and the frequency of
such mergers out to extragalactic distances. Based on the expected
sensitivity of LIGO I and II, a NS--NS inspiral could be detected out to
20\,Mpc and 350\,Mpc, respectively (\cite{G99}1999). Predictions for the
expected detection rates can be made based on estimates of the Galactic
coalescence rate and its extrapolation to these maximum distances of
interest. Such Galactic estimates have been obtained in two different ways
so far: empirically, based on the observed NS--NS sample, and purely
theoretically, based on our understanding of NS--NS formation.

Binary neutron stars can be discovered in radio pulsar surveys, if one of
the two neutron stars emits radio pulses.  In the last two decades, four
NS--NS binaries in addition to PSR~B1913+16 have been discovered by
sensitive pulsar searches: PSR~B1534+12 (\cite{W91}1991), PSR~B2127+11C
(\cite{Pr91}1991), PSR~J1518+4904 (\cite{N96}1996), and PSR~J1811-1736
(\cite{L00}2000). Of these four,
only the first two are {\em coalescing} binaries, along with PSR~B1913+16,
i.e., have tight enough orbits so that the two neutron stars will coalesce
within $10^{10}$\,yr. One other binary pulsar, PSR~B2303+46
(\cite{LB90}1990), was classified as a NS--NS binary for many years, until
it was optically identified (\cite{KK99} 1999) and is now thought to be a
binary pulsar with a massive white dwarf companion. Recently, another
binary pulsar, PSR~J1141-6545, with a relatively short coalescence time
was discovered by the ongoing Parkes multibeam pulsar survey
(\cite{L00}2000) and was also initially considered to be a NS--NS
candidate. However, the low measured total mass of the system points to a
white dwarf companion to the pulsar rather than a neutron star
(\cite{Kas00}2000). In the analysis presented in this paper, we include
the two established NS--NS binaries with short merger times found in the
Galactic field, i.e., PSR~B1913+16 and PSR~B1534+12. We do not include
PSR~B2127+11C, which is found in a globular cluster, for two main reasons:
(i) its association with globular clusters implies very different
formation history (dominated by dynamical interactions) and detection
selection effects, and therefore our analysis cannot be applied to such
systems, and (ii) based on the globular cluster space density, the
contributions of cluster systems to the coalescence rate has been
estimated by \cite{P91}(1991) to be very small (by more than a factor of
10).

These discoveries have contributed to our knowledge of the properties of
NS--NS binaries and allow empirical estimates of the coalescence rate,
based on a quantitative analysis of the selection effects relevant to
pulsar surveys (e.g., \cite{N91}1991; \cite{P91}1991). Also, over the
years, a theoretical understanding of the NS--NS formation process has
developed and estimates of their birth rate have been obtained based on
theoretical calculations of binary evolution (e.g., \cite{L97}1997;
\cite{FBB98}1998; \cite{PZ98}1998;  \cite{BB98}1998; \cite{FWH99}1999;
\cite{B99}1999; \cite{G01}2001). Still, however, serious
uncertainties remain that hamper settling the question of whether close
NS--NS systems are formed at adequately high rates to provide a reasonable
LIGO detection rate (at least a few events per year). At present,
theoretical estimates cover a wide range of values (three to four orders
of magnitude; see \cite{K00a}2000a) and appear to have a rather limited
predictive power\footnote{However, taking into account multiple
observational constraints on the absolute calibration of population
synthesis models can significantly improve these estimates. See, for
example, \cite{BK00} 2000.}.

Neutron star coalescence has also been discussed as a possible central
engine of gamma-ray bursts (e.g., \cite{P86}1986; \cite{E89}1989;
\cite{NPP92}1992) and a production site of r-process elements
(\cite{E89}1989; \cite{R99}1999). Among other issues, accurate estimates
of merger frequencies could be useful in examining these proposed
associations and possibly constraining the degree of beaming in gamma-ray
burst emission (e.g., \cite{Bel99}1999).

In this paper we focus on estimates of the Galactic birth rate of
coalescing NS--NS binaries based on the current observed sample. In \S\,2,
we give a brief overview of such empirical estimates and the main steps
involved in the calculations. In \S\,3, we describe our derivation of the
NS--NS coalescence rate in our Galaxy, addressing each of the
uncertainties involved: NS--NS scale height in the Galaxy, system
lifetimes, corrections for the faint end of the radio pulsar luminosity
function, beaming, and pulse smearing due to orbital acceleration.  In
\S\,4, we extrapolate our Galactic estimate to extragalactic distances
relevant to LIGO I and II. In \S\,6, we present our conclusions on the
NS--NS coalescence rate, the associated uncertainty, and the expected LIGO
detection rates.  We also compare our results to the theoretical estimates
derived based on binary evolution calculations.

\section{PREVIOUS ESTIMATES OF THE GALACTIC NS--NS COALESCENCE RATE}

Estimates of the NS--NS coalescence rate in our Galaxy can be obtained
using the observed properties of the radio pulsars in such binaries and
the characteristics of radio pulsar surveys. Based on these two elements,
a model can be constructed accounting for pulsar selection effects and the
detectability of the observed pulsars throughout the Galaxy. For each
pulsar observed in a coalescing NS--NS system, a {\em scale factor},
${\cal S}$, can be calculated (e.g., \cite{N87}1987). It is defined as the
inverse of the fraction of the Galactic volume (weighted by the
radio-pulsar spatial distribution in the Galaxy), within which pulsars
with properties identical to those of the observed pulsar could be
detected by any of the pulsar surveys completed so far,
 \begin{equation}
 {\cal S}~=~\frac{\int_{V_G}~F_p\left(R,Z\right)~dV}
{\int_{V_D}~F_p\left(R,Z\right)~dV}, 
 \end{equation}
 where $F_p$ describes the radial ($R$) and vertical ($Z$) distribution of
pulsars in the Galaxy ($F_p$ is usually assumed to be axisymmetric and
separable in $R$ and $Z$), $V_G$ is the Galactic volume, and $V_D$ is the
volume within which each observed pulsar could be detected by the pulsar
surveys.  The scale factor ${\cal S}$ is a measure of how many more
pulsars like those already detected in coalescing NS--NS binaries exist in
the Galaxy. The coalescence rate is then estimated using these scale
factors divided by estimates for the lifetime of each radio pulsar summed
up over all detected coalescing NS--NS. This estimate can be further
corrected for a fraction of undetected pulsars either because the pulsar
beam does not intersect our line of sight (beaming fraction), or because
they are too faint to be detected even by the most sensitive surveys
conducted so far, or because the binary orbital periods are so short that
the pulses are smeared due to orbital acceleration.

This method was first applied by \cite{N91}(1991) and \cite{P91}(1991) to
obtain empirical estimates of the NS--NS coalescence rate in the Galaxy.
\cite{N91}(1991) adopted a Gaussian form for $F_p$ ($\propto \exp
[-(R/8{\rm kpc})^2 - (Z/Z_0)^2/2]$, where $Z_0$ is the vertical scale
height), and \cite{P91}(1991) adopted a constant pulsar density in a
cylinder of radius $R_0=12$\,pc and half-height $Z_0$. Both groups
considered the same pulsar surveys and made similar assumptions about the
lifetimes of PSR~B1913+16 and PSR~B1534+12. They assumed that the lifetime
of a pulsar binary is equal to the sum of the pulsar characteristic age
($\tau_c=P/2\dot{P}$) and the binary merger time $T_{m}$, i.e., the time
it will take for the binary to coalesce through gravitational radiation
emission. Without any further corrections for beaming, the faint end of
the pulsar luminosity function, or pulse smearing due to orbital
acceleration, both studies obtained an estimate of the Galactic NS--NS
coalescence rate equal to $10^{-6}$\,yr$^{-1}$ assuming $Z_0=1$\,kpc.

These early estimates have been subsequently revised as more of the
Galactic volume was searched for pulsars.  \cite{CL95}(1995) included more
pulsar surveys, which did not lead to new NS--NS discoveries. However,
they did not include the Green Bank Northern Sky Survey that discovered
PSR~J1518+4904.  They assumed the same lifetimes as \cite{N91}(1991) and
\cite{P91}(1991), and investigated in more detail the dependence of the
derived rate on the assumed radial and vertical pulsar distribution in the
Galaxy. They found that the scale factors decrease by a factor of two if a
radial distribution with a strong deficit of pulsars in the inner Galactic
region (\cite{J94}1994) is assumed. The dependence on the scale height was
found to be linear assuming an exponential vertical distribution ($\propto
\exp(-\vert Z\vert /Z_0)$. For $Z_0=0.5$\,kpc, a radial distribution
similar to that assumed by \cite{N91}(1991), and without any further
corrections, they obtain a rate estimate of $\simeq
2\times10^{-7}$\,yr$^{-1}$. The primary reasons for this significant
reduction compared to the earlier estimates were (i) the use of a
different pulsar distance model, which leads to a higher luminosity for
PSR~B1534+12, and (ii) the lack of any new discoveries of NS--NS binaries
with the additional pulsar surveys. \cite{CL95}(1995) also included a
beaming correction of a factor of $3$ and a 10-fold correction for the
undetected pulsars at the faint end of the luminosity function
(extrapolation down to 1\,mJy\,kpc$^2$) raising the estimate to $\simeq
6\times10^{-6}$\,yr$^{-1}$.

Van den Heuvel \& Lorimer (1996) reconsidered the NS--NS lifetimes and
argued that the binary merger time should be replaced by the time in
which the luminosity of the observed pulsars drops below a detection
threshold (when the magnetic dipole energy loss rate reaches that of a
typical normal pulsar after 10\,Myr). This strongly model--dependent
modification was applied {\em along with} the 10--fold correction for the
faint pulsars introduced by \cite{CL95}(1995). Based on this
\cite{vdHL96}(1996) concluded that the NS--NS coalescence rate is raised
by a factor of 2.7 compared to the \cite{CL95}(1995) result. More
recently, measurements of relativistic orbital parameters (\cite{S98}1998)
have provided us with a precise distance measurement for PSR~B1534+12,
which led to a downward revision of the \cite{vdHL96}(1996) rate to
$\simeq 6.5-8.5\times 10^{-6}$\,yr$^{-1}$.

\cite{A98}(1998) examined once again the issue of the NS--NS lifetimes.
They argued that a better estimate of the present age can be obtained
based on the pulsar spin--down history, assuming that their life started
at the spin--up line in the period--period derivative plane for pulsars
(uncertainties on the definition of the spin--up line lead to age
variations smaller than a factor of 2). Concerning the remaining lifetime
of the binaries, they adopted the ``luminosity-evolution'' lifetime
suggested by \cite{vdHL96}(1996), but also considered pulsar selection
effects against short-period binaries, because of pulse smearing (it
becomes important only for PSR~B1913+16). Using the scale factors derived
by \cite{CL95}(1995), they revised the rate to
$2.7\times10^{-7}$\,yr$^{-1}$, and to $8\times10^{-6}$\,yr$^{-1}$ when
they included the same corrections for beaming and faint pulsars. In
addition, \cite{A98}(1998) obtained {\em another} rate estimate based on
(i) a revised (increased by more than a factor of 3) total Galactic volume
$V_G$, (ii) a revised detection volume $V_D$ integrating out to maximum
detection distances and over the pulsar luminosity function (in this way,
a correction for the faint end of the luminosity function was incorporated
to some extent but no weighting based on the Galactic distribution of
pulsars was applied), and (iii) an average lifetime for all NS--NS
systems, since no individual scale factors were calculated. They obtained
a much lower estimate of $\simeq 2\times10^{-7}$\,yr$^{-1}$ (this does not
include any beaming correction).

\cite{E99}(1999) followed a rather different approach and argued that the
calculations of the detection volume and the pulsar lifetimes cannot be
performed separately because pulsar luminosity evolution affects both
simultaneously.  Instead, for each of the two coalescing NS--NS systems,
they calculated an average {\em visibility factor}. This factor accounts
for the motion of such systems in the Galaxy, given a range of initial
velocities and taking into account the selection effects associated with
all pulsar surveys to date. The sum of the reciprocals of these visibility
factors is an estimate of the coalescence rate. Preliminary results yield
an estimate of $7\times10^{-8}$\,yr$^{-1}$ (without any beaming
correction).

\section{GALACTIC NS--NS COALESCENCE RATE AND UNCERTAINTIES}

In what follows we present our derivation of the Galactic NS--NS
coalescence rate addressing each one of the elements that enter the
calculation. For some of these elements, we use the results of earlier
studies but we examine in more detail the associated uncertainties and the
sensitivity of our estimate to various factors.

\subsection{Scale Factors}

The calculation of scale factors depends on the pulsar surveys completed
so far and the assumed spatial distribution of NS--NS binaries.
\cite{CL95}(1995) have considered all the pulsar surveys to date except
for the ongoing Parkes Multibeam survey (\cite{L00}2000). They also
studied the dependence of the scale factors on the assumed radial
distribution. Combined with the results obtained by \cite{N91}(1991) and
\cite{P91}(1991), it becomes evident that the scale factors are not highly
sensitive to such variations, unless the radial distribution is strongly
modified (large deficit of pulsars in the center of the Galaxy, see
\cite{J94}1994). Here, we use the scale--factor results obtained by
\cite{CL95}(1995), for a Gaussian radial distribution with a scale length
$R_0=4.8$\,kpc (higher by a factor of $\simeq 2$ -- within their
statistical errors -- relative to the results obtained when the
\cite{J94}1994 radial distribution is assumed; see Table 1 and Figure 2 in
\cite{CL95}1995) and an assumed scale height $Z_0=0.5$\,kpc. We also take
into account the significant reduction of the scale factor for
PSR~B1534+12, caused by the relativistic measurement of its distance
(\cite{S98}1998). Depending on the scale height of the NS--NS population
this reduction factor could be as low as 2.5 (if $Z_0 < 1$\,kpc) and as
high as 4 (if $Z_0 > 1$\,kpc). Since \cite{CL95}(1995) derived the scale
factors assuming $Z_0=0.5$\,kpc, we apply a correction factor of 2.5. So,
the scale factors we use in our rate derivation are ${\cal S}=40$ for
PSR~B1913+16 and ${\cal S}=130$ for PSR~B1534+12. Statistical errors
associated with these values are $\simeq 5-6$\%. These scale factors will
be further modified by our calculation of $Z_0$ based on the linear
dependence derived by \cite{CL95}(1995).

\subsection{Vertical Scale Height}

\cite{N91}(1991) found that the coalescence rate is roughly linearly
proportional to the scale height, $Z_0$, of the Galactic NS--NS
population. This dependence originates mainly from the $Z_0$ dependence of
the total Galactic volume weighted by the NS--NS spatial distribution (see
equation [1]) . However, due to the spatial distribution weighting, the
scale height enters the calculation of the detection volume as well.
\cite{CL95}(1995) examined this dependence in more detail and found that
the scale factor (summed up for the three binaries PSR~B1913+16,
PSR~B1534+12, and PSR~B2303+46) has the following linear dependence:
${\cal S}_{\rm total}\propto 0.1(Z_0/{\rm kpc})+0.19$, assuming an
exponential vertical distribution: $\propto \exp(-\vert Z\vert /Z_0)$. The
value of $Z_0$ is usually assumed to lie in the range $0.5-1$\,kpc.
However, we undertake a more detailed study and we calculate it based on
realistic models for the motion of coalescing NS--NS binaries in the
Galaxy. In what follows we describe our approach in detail.

\subsubsection{Physical Model and Numerical Method} 

The formation history of coalescing NS--NS binaries involves a number of
evolutionary stages including two supernova explosions. The associated
mass loss and birth kicks imparted to the neutron stars affect the
center--of--mass velocities ($V_{\rm CM}$) of the binaries. The
post--supernova velocities are closely related to the relative orbital
velocity ($V_r$) in the pre--supernova orbit and a strict upper limit of
$2V_r$ to $V_{\rm CM}$ can be derived (e.g., \cite{BP95}1995;
\cite{K96}1996). The binary orbits of NS--NS progenitors before the {\em
first} supernova (SN) explosion are so wide that the typical
post--explosion center--of--mass velocities are $V_{\rm CM}\lesssim
50-75$\,km\,s$^{-1}$, i.e., significantly lower than the typical
velocities associated with Galactic rotation.  However, just prior to the
{\em second} SN explosion, progenitors of coalescing NS--NS are expected
to have orbits tight enough that typical post--explosion velocities reach
or even exceed $500$\,km\,s$^{-1}$. Such high velocities alter
significantly the kinematic evolution of the NS--NS population and lead to
vertical scale heights much larger than that of their initial progenitors
(the typical scale height of massive stars is $\sim 50-75$\,pc).

In what follows we consider NS--NS progenitors just before the {\em
second} SN explosion. According to all NS--NS formation mechanisms
discussed in the literature (see \cite{FK97}1997 for a brief overview),
the binaries at this stage consist of the first NS and a helium-rich
companion (the core of the original secondary, i.e., less massive star, in
the binary) in circular orbits. In fact, (i) circular
pre--SN\footnote{From here and on, SN refers specifically to the second
supernova explosion in the formation history of NS--NS binaries} orbits
with sizes $\lesssim 50-100$\,R$_\odot$, and (ii) helium-star companions
with masses in the range $3-10$\,M$_\odot$ are necessary for the formation
of tight, {\em coalescing} NS--NS binaries (see \cite{FK97}1997;
\cite{FWH99}1999; \cite{W00}2000; Fryer 1999, private communication). Such
progenitor properties are achieved through common--envelope evolution,
which has to occur some time between the two supernova explosions. We
assume that isotropic birth kicks of a certain magnitude distribution are
imparted to the newborn neutron stars and we calculate the $V_{\rm CM}$
distribution for the subset of post-SN systems that (i) remain bound after
the explosion, and (ii) have binary properties (post--SN orbital
separations and eccentricities)  such that their merger times do not
exceed $10^{10}$\,yr.

For given values of the helium--star mass $M_0$, the pre-SN orbital
separation $A_0$, and the isotropic kick magnitude $V_k$, we can use
conservation laws of energy and angular momentum for the system and derive
expressions for the post--SN orbital semi--major axis $A$ and eccentricity
$e$. The center-of-mass velocity $V_{\rm CM}$ after the explosion can also
be calculated (see \cite{H83}1983; \cite{BP95}1995; \cite{KL00}2000):
 \begin{equation}
 A~=~\frac{\beta\,A_0}{2\beta-u_k^2\,\sin^2\theta
-(u_k\cos\theta+1)^2},
 \end{equation}
 \begin{equation}
 1-e^2~=~\frac{1}{\beta^2}\left[u_k^2\,\sin^2\theta\,\cos^2\phi+
(u_k\cos\theta+1)^2\right]\,\left[2\beta-u_k^2\,\sin^2\theta
-(u_k\cos\theta+1)^2\right], 
 \end{equation}
 \begin{equation}
 \left(\frac{V_{\rm CM}}{V_{\rm orb}}\right)^2~=~\frac{M_{\rm NS}
(M_0-M_{\rm NS})}{2(M_0+M_{\rm
NS})^2}~\frac{A_0}{A}~+~\frac{M_0}{2(M_0+M_{\rm
NS})}\left(\frac{V_k}{V_{\rm orb}}\right)^2~+~
\frac{(M_0-M_{\rm NS})(M_0-2M_{\rm NS})}{2(M_0+M_{\rm NS})^2}. 
 \end{equation}
 In the above equation $V_{\rm orb}\equiv [G(M_{\rm NS}+M_0)/A_0]^{1/2}$
is the pre--SN relative orbital velocity, $\beta\equiv (M_{\rm NS}+M_{\rm
NS})/(M_{\rm NS}+M_0)$, and angles $\theta$ and $\phi$ describe the
direction of the kick: $\theta$ is the polar angle from the pre-SN orbital
velocity vector of the exploding He-star and ranges from $0-\pi$ (at
$\theta=0, \vec{V}_k$ and $\vec{V}_{\rm orb}$ are aligned); $\phi$ is the
azimuthal angle in the plane perpendicular to $\vec{V}_{\rm orb}$ (i.e.,
$\theta=\pi /2$) and ranges from $0-2\pi$ (at $\theta=\pi /2$ and $\phi=0$
or $\phi=\pi$, the kick component points along or opposite to the angular
momentum axis of the pre--SN orbital plane, respectively; see Figure 1 in
\cite{K00b}2000b). For a given pair of $A$ and $e$, the {\em merger time
scale} $T_{\rm m}$ is given by (\cite{PM63}1963; \cite{ST83}1983) as
 \begin{equation}
 T_{\rm m}\equiv \left(\frac{d\ln E_{\rm
orb}}{dt}\right)^{-1}\simeq~300\,{\rm Myr}\,\left(\frac{M_{\rm
NS}}{M_\odot}\right)^{-3}\,\left(\frac{A}{R_\odot}\right)^4\,
\left(1-e^2\right)^{7/2}\,
\left(1+\frac{73}{24}e^2+\frac{37}{96}e^4\right)^{-1}. 
 \end{equation}
 We note that this is just the time scale associated with the orbital
energy loss rate due to gravitational radiation, and not the time interval
over which a binary of given $A$ and $e$ will merge (this can be
calculated by solving the coupled equations for the rate of change of both
orbital separation and eccentricity; \cite{PM63}1963). Comparison of these
two quantities shows that, because orbital shrinkage rapidly accelerates
with time, the merger time scale typically {\em over}estimates the actual
merger time by factors of $\lesssim 4$ (for circular orbits, the
difference is exactly a factor of 4; in the cases of PSR~B1913+16 and
PSR~B1534+12, the factors are 1.87 and 3.3, respectively). In what follows
we use the merger time scale, $T_{\rm m}$. However, we examine the effect
of this difference on the derived scale heights. We find that if we
decrease the calculated $T_{\rm m}$ by a uniform factor of 3, the scale
height decreases (as expected qualitatively) by less than 2.5\%. 

We first derive the distribution function of $A$ and $e$ from the
distribution of the direction angles, $\theta$ and $\phi$, for an
isotropic kick with a simple Jacobian transformation
 \begin{eqnarray}
F(A,e)&=&F(\theta,\phi)J\left(\frac{\theta,\phi}{A,e}\right)~=~
\frac{\sin\theta}{2}\,\frac{1}{2\pi}~
J\left(\frac{\theta,\phi}{A,e}\right) \nonumber \\
 &=&\frac{\beta^2 e V_{\rm orb}}{2\pi V_k A}\left[\beta(1-e^2)A/A_0-
\left(\frac{2\beta-\beta A_0/A -(V_k/V_{\rm orb})^2
-1}{2}+1\right)^2\right]^{-1/2} \nonumber \\
 &&\times \left[2\beta-\beta A_0/A-\beta(1-e^2)A/A_0\right]^{-1/2}. 
 \end{eqnarray}
 With a similar transformation between $A$ and $V_{\rm CM}$ and
integration over $e$ we can obtain numerically
 \begin{equation}
 F(V_{\rm CM})~=~\int_0^1~F(A,e)~J\left(\frac{A,e}{V_{\rm
CM},e}\right)~de. 
 \end{equation}
 However, the dynamical evolution of the whole population in the Galactic
potential depends not only on the initial velocity distribution but also
the pulsar lifetimes. Any strong correlation between the two quantities
can affect the final vertical distribution of the population. In fact,
from studies of the effects of SN kicks on binary properties (e.g.,
\cite{K96}1996), we know that post-SN binaries with tight or significantly
eccentric orbits tend to acquire large center--of--mass velocities. Such
tight and eccentric orbits, however, will typically have short merger time
scales. Consequently, systems with large initial velocities, which could
drive the population to high vertical scale heights, typically have short
timescales, which counteracts their expansion in the Galaxy. To include
all these effects rigorously in our calculation, we derive the
two--dimensional distribution function of $V_{\rm CM}$ and $T_{\rm m}$,
for given $V_k$, $M_0$, $A_0$:
 \begin{eqnarray}
 F\left(V_{\rm CM},T_{\rm m};
V_k,M_0,A_0\right)&=&F(A,e)~J\left(\frac{A,e}{V_{\rm CM},T_{\rm
m}}\right) \nonumber \\
 &=&\frac{4A^2V_{\rm CM}}{A_0V^2_{\rm orb}T_{\rm m}}~\frac{(M_0+M_{\rm
NS})^2}{M_{\rm NS}(M_0-M_{\rm NS})}~F\left[A(V_{\rm CM}),e(V_{\rm
CM},T_{\rm m})\right] \nonumber \\
 & & \times
\left[\left(2\frac{73}{24}e+4\frac{37}{96}e^3\right)\left(1+
\frac{73}{24}e^2+\frac{37}{96}e^4\right)^{-1}+\frac{7e}{1-e^2}\right]^{-1}.
 \end{eqnarray}
 All the derivatives involved in the Jacobian transformations appearing in
equations (6)--(8) are calculated analytically using equations (2)--(5).
From equation (4), we obtain $A$ as a function of $V_{\rm CM}$
analytically, and from equation (5), we obtain that of $e$ as a function
of $T_{\rm m}$ numerically. The next step is to convolve the above
distributions with a kick magnitude distribution as well as the
distribution functions of $M_0$ and $A_0$, which we assume to be
separable:
 \begin{eqnarray}
 F\left(V_{\rm CM},T_{\rm
m}\right)&=&\int_{M_0}\int_{V_k}\int_{A_0}\,F\left(V_{\rm CM},T_{\rm
m}; V_k,M_0,A_0\right)\,\times \nonumber \\ 
 & & \hspace{2.cm}
f_{A_0}\left(A_0\right)\,f_{M_0}\left(M_0\right)\,f_{V_k}\left(V_k\right)
\,dA_0dV_kdM_0.
 \end{eqnarray}  
 The ranges of integration over $A_0$ are dependent on $V_k$ and $M_0$ and
are determined by the requirement that the post-SN binaries are {\em
coalescing}. The range of integration over $M_0$ is restricted to those
helium star progenitors that are expected to end their lives as neutron
stars (instead of black holes), i.e., $2-10$\,M$_\odot$ (\cite{F99}1999
and private communication). It turns out that ranges of interest for $A_0$
and $M_0$ are narrow enough that our final results are not at all
sensitive to the details of the chosen distribution functions. In what
follows, we adopt an $A_0$--distribution flat in $\log A_0$ and a
power--law function for $f_{M_0}(M_0)\,dM_0\,\propto
\,M_0^{-\alpha}\,dM_0$ with $\alpha$ in the range $1-3$ (these choices are
motivated by typical assumptions about the characteristics of the
primordial binary population, e.g., \cite{KW98}1998; \cite{PZ98}1998). The
two--dimensional distribution can be also integrated over $V_{\rm CM}$ or
$T_{\rm m}$ to obtain one--dimensional distributions (as described in the
next section).

We consider a population of NS--NS progenitors distributed in the Galaxy
with a Gaussian radial distribution in azimuthal symmetry. The population
is assumed for simplicity to lie in the Galactic plane. This assumption is
motivated by the very small (much smaller than the expected scale height
of NS--NS binaries) scale height of the Galactic massive star population
and the small (smaller than typical Galactic rotational velocities)
center--of--mass velocities after the {\em first} supernova explosion. We
have checked and confirmed the validity of this assumption using our
numerical results and studying models where the initial population is
assumed to lie at $Z=\pm 100$\,pc. The resulting distributions of binaries
in the vertical direction are identical to an accuracy of $\sim 0.1$\% or
better. For the orbit calculations, we use the Galactic gravitational
potential of \cite{KG89}(1989).  We make sure that the Galactic rotation
curve used in the calculation is consistent with the chosen potential.

We follow the dynamical evolution of the NS--NS population in the
five--dimensional phase space of two spatial dimensions (radial $R$ and
vertical $Z$ positions) and three velocity components ($u_\Phi$, $u_R$,
$u_Z$), all defined in a reference frame of cylindrical coordinates
centered on the Galactic center. The distribution function that describes
the binary population at $t=0$ is actually six--dimensional because it
includes the merger time scale ($T_{\rm m}$), although it does not evolve
with time:
 \begin{equation}
 F_0\left(R,Z,u_\Phi,u_R,u_Z,T_{\rm m}\right)~=~f_R(R)\,\delta(Z)\,
f_u(u_\Phi,u_R,u_Z,T_{\rm m}), 
 \end{equation} 
 i.e., we assume separation of variables for the spatial components and
that the initial $Z$--distribution is a delta function in the Galactic
plane ($Z=0$). The initial velocity distribution $f_u$ is calculated using
the distributions of post--SN center--of--mass velocities for coalescing
NS--NS binaries $F(V_{\rm CM},T_{\rm m})$ and assuming that they are
imparted to the binaries {\em isotropically} with respect to their
Galactic rotation velocities (the latter depend on their radial position
and the assumed Galactic potential).

We first set up a grid of initial phase--space positions and we calculate
the Galactic orbits in the chosen five\footnote{Note that azimuthal
symmetry is conserved since both the gravitational potential and the
initial spatial distribution are axisymmetric}--dimensional phase space as
a function of time, for a maximum time of 10$^{10}$\,yr. We have tested
numerically the initial grid and adopted one of high enough density so
that numerical artifacts in the analysis are avoided. In the present
study, we are interested in the vertical distribution of the coalescing
NS--NS binaries. For a given Galactic gravitational potential, the
vertical distance $Z$ is a unique function of the time $t$, for which they
have been moving in the Galaxy, and the initial phase space positions
$Z^\prime=Z(t,R,u_\Phi,u_R,u_Z)$. Using exactly this functional
dependence, which is calculated numerically by calculating the orbits in
phase space, we can calculate the $Z$--distribution at a given time $t$
(for a grid of $t$ values up to $10^{10}$\,yr). The calculation involves a
one-dimensional Jacobian transformation and a multiple numerical
integration over the initial phase space variables:
 \begin{equation}
 F_t\left(Z^\prime,T_{\rm
m}\right)~=~\int_R\,\int_{u_\Phi}\,\int_{u_R}\,f_R(R)\,f_u(u_\Phi,u_R,u_Z,
T_{\rm m})\,\left[\left(\frac{\partial Z^\prime(t)}{\partial
u_Z}\right)_{R,u_\Phi,u_R}\right]^{-1}~du_R\,du_\Phi\,dR.   
 \end{equation} 
 The integration limits are chosen based on the shape of the distribution
functions $f_R$ and $f_u$ and the range of values they cover (for $f_u$,
we take into account the $V_{\rm CM}$ distribution calculated for each
model and the contribution of Galactic rotation velocities as a function
of $R$). We have performed numerical tests to examine the effects of the
chosen limits and we make sure that their effect on the integral lies well
below 1\%.

We are interested in the $Z$--distribution of coalescing NS--NS binaries at present,
calculated taking into account (i) the star formation history of the Galaxy and (ii)
the distribution of merger time scales, which can be significantly shorter than
$10^{10}$\,yr. Assuming a constant (per unit time) star formation rate, $C_{\rm
SFR}$, over the age of the Galaxy, the distribution function describing the NS--NS
population, i.e., number of systems per unit $Z^\prime$, per unit merger time, and
per unit formation time over the age of the Galaxy is given by:
$F_t\left(Z^\prime,T_{\rm m}\right)\times C_{\rm SFR}$. To obtain the
$Z$--distribution of coalescing NS--NS binaries at present, we have to integration
the latter distribution function over all merger times, $T_m$, and all formation
times, $T_F$. Since $t$ is the time since the second supernova explosion, their
formation time is: $T_F=10^{10}-t$. The range of integration for $T_F$ is determined
by the fact that $t$ cannot exceed the associated merger time scale ($t\leq T_{\rm
m}$). To obtain the $Z$-distribution at present, we then have to integrate over all
$T_{\rm m}$, yielding:
 \begin{eqnarray}
 F\left(Z^\prime\right)&=&C_{\rm SFR}\,\int_{T_{\rm m}=0}^{T_{\rm
m}=10^{10}}\,\int_{T_F=10^{10}-T_{\rm m}}^{T_F=10^{10}}\,
F_t\left(Z^\prime,T_{\rm m}\right)\,dT_F\,dT_{\rm m} \nonumber \\
 &\propto& \int_{T_{\rm m}=0}^{T_{\rm
m}=10^{10}}\,\int_{t=0}^{t=T_{\rm m}}\,
F_t\left(Z^\prime,T_{\rm m}\right)\,dt\,dT_{\rm m}.  
 \end{eqnarray}
 Since the goal of this calculation is to obtain the scale
height $Z_0$, we do not have to worry about the absolute normalization of
the distribution function (e.g., the value of $C_{\rm SFR}$). 

We wish to note that the computational method of
distribution--function evolution in an appropriate phase space is very
different from Monte Carlo methods that are more widely used for these
types of problems (binary population synthesis and dynamical evolution of
populations in a fixed potential). Our choice of methods has been driven
by their high numerical accuracy at a relatively low computational cost. A
general discussion and comparison can be found in \cite{KB00}(2000).
Specific to the problem of dynamical evolution, an additional advantage of
the method is that it allows us to calculate the needed grid of orbits
only once (this is the most computationally demanding part of the
calculation) and then use it in the parameter study. 

\subsubsection{Results}

We calculate the distribution of coalescing NS--NS binaries in birth
center--of--mass velocities, merger time scales, and vertical distance
from the Galactic plane, for different kick--magnitude distributions: a
set of Maxwellian distributions with
$\sigma=100,200,300,400$\,km\,s$^{-1}$ and a ``Paczynski--like''
(\cite{P90}1990)  distribution with a large fraction of small--magnitude
kicks: $f_k(V_k)\propto (1+u^2)^{-1}$, where $u=V_k/600$\,km\,s$^{-1}$.
These distributions are consistent with studies of the radio--pulsar
population (e.g., \cite{CC97}1997; \cite{H97}1997; \cite{FBB98}1998), but
also cover an adequately large range of average kick magnitudes.

A set of distribution functions (normalized to unity) of $V_{\rm CM}$ are
shown in Figure 1, for a specific helium--star mass ($M_0=4$\,M$_\odot$,
as an example). It is evident that birth $V_{\rm CM}$ tends to increase
with average kick magnitude initially ($\sigma=100,200$\,km\,s$^{-1}$),
but becomes less and less sensitive with higher kicks
($\sigma=300,400$\,km\,s$^{-1}$). The basic reason for this behavior is
that we are examining a certain subgroup of post--SN systems that
satisfies two specific constraints: they are bound {\em and} have merger
times shorter than $10^{10}$\,yr. These two constraints act as a {\em
filter} on the NS--NS population and all their properties: As $\sigma$
increases initially, the velocities of this subgroup increase as well.
However, as kicks become higher, systems with even higher center--of--mass
velocities no longer satisfy these two constraints and are ``filtered out''
of the population of interest to us. This behavior is in excellent
agreement with the basic understanding of the effects of kicks on binary
populations (e.g., \cite{K96}1996). We note that the ``Paczynski--like''
distribution does not lead to significantly different results from the
Maxwellian distributions, even though it includes a significant component
of low--velocity kicks. The reason is again this ``filtering''
effect: formation of {\em coalescing} NS--NS requires some minimum kick
(\cite{FK97}1997; \cite{W00}2000)  and the existence of small kicks does
not affect the properties of the final population\footnote{The formation
rate is of course greatly affected depending on the fraction of kick
magnitudes that are in the ``favored window'' for the formation of tight
NS--NS binaries. This sensitivity of formation is clearly evident in our
results but does not affect our calculation of the scale height $Z_0$,
which is independent of the overall rate. For this reason, we show only
the normalized distributions of various parameters.}. 

We should also mention that higher velocities are acquired by systems with
more massive helium--star progenitors (we cover a range of
$2-10$\,M$_\odot$ determined by the requirement that a NS is formed).
However, such higher masses are disfavored by the mass function of
progenitors (inverse power--law) and therefore their contribution in the
final $Z$--distributions is limited. The range in $M_0$ is narrow enough
that the details of the mass distribution become unimportant. This is
clearly indicated by the insensitivity ($\lesssim 1$\%) of our results on
the specific choice of the power--law index over a wide range of values
($1-3$).

In Figure 2 a set of distribution functions (normalized to unity) of
$T_{\rm m}$ are plotted, for different kick--magnitude distributions and
for a specific helium--star mass. As in the case of velocities, the
dependence of the distributions becomes much weaker as average kick
magnitudes increase. It is interesting to note that the cases with
typically higher $V_{\rm CM}$ also lead to typically shorter merger time
scales. This correlation originates from the fact that velocities are
higher for tighter binaries, which also experience faster orbital decay
(and merging), and as a result, the expansion of the coalescing NS--NS
population from their birth places is somewhat limited.

Our final $Z$--distributions are shown in Figure 3. It is evident that --
not surprisingly -- the actual shape of these distributions is not a
perfect exponential ($\exp(-\vert Z\vert /Z_0$) as assumed in studies of
pulsar selection effects (this is true even if we plot the distributions
as a function of $Z^2$, for the form $\exp(-Z^2/2Z_0^2$). We explore the
effect of this difference on the scale factors by adopting two different
ways of obtaining a scale height for the NS--NS population. (i) For an
exponential distribution, the fraction of the population with vertical
distances smaller than $Z_0$ is equal to $63.2\%$. For our calculated
$Z$--distributions, we adopt a scale height so that $63.2\%$ of the
population lies at smaller vertical distances.  For the four different
kick--magnitude models, the derived scale heights lie in the range
$0.8-1.7$\,kpc. Based on the $Z_0$ scaling derived by \cite{CL95}(1995),
this $Z_0$--range leads to scale factors ${\cal S}$ in the ranges $45-60$
for PSR~B1913+16 and $145-195$ for PSR~B1534+12. (ii) We note that
although the derived $Z$--distributions cannot be well described by a
single exponential, they {\em can} be fitted rather well by exponentials
if they are divided into two segments. We choose the ``break'' point to be
at $Z=0.5$\,kpc and we calculate two scale heights: 0.3\,kpc for
$Z<0.5$\,kpc (same for all kick--velocity models) and $1.5-2.5$\,kpc for
$Z>0.5$\,kpc (for the various kick--velocity models). We then use the
scale heights and relative fractions of the two sub--populations and find
the scale factors ${\cal S}$ in the ranges $46-61$ for PSR~B1913+16 and
$148-197$ for PSR~B1534+12.

We find the agreement in the scale factors from these two different types
of analysis quite encouraging. Accordingly, we adopt the following
conservative ranges for the scale factors: $45-60$ for PSR~B1913+16 and
$145-200$ for PSR~B1534+12.

\subsection{NS--NS Lifetimes}

To obtain a coalescence rate for NS--NS binaries, estimates of their
lifetimes are necessary. In particular, we are interested in the
``observable'' lifetimes of the systems, i.e., the time intervals during
which the observed NS--NS binaries are detectable.  They are the sum of
the time since their formation (current age) and of their remaining
lifetime.

The characteristic age $\tau_c\equiv P/2\dot{P}$, where $P$ and $\dot{P}$
are the spin period and its derivative, is often used as a measure of the
current age of a radio pulsar. This is true assuming that magnetic dipole
radiation roughly describes the pulsar emission, the magnetic braking
index is equal to $3$, and the initial spin period was much smaller
than the current period. It is not always clear that these assumptions are
valid, but in many cases the characteristic age serves as a good
approximation of the true pulsar age, especially in the absence of any
other information. The characteristic ages for PSR~B1913+16 and
PSR~B1534+12 are $1.1\times 10^8$\,yr and $2.5\times 10^8$\,yr,
respectively. However, the radio pulsars found in coalescing NS--NS
binaries appear to have been recycled. This implies that the current spin
periods are not much different than the initial periods at the end of the
recycling phase. \cite{A98}(1998) suggested that an alternative and
possibly better way to obtain age estimates is to consider the spin
history of these pulsars. They calculated the initial spin, $P_0$, as
indicated by the intersection of the magnetic dipole spin-down line and
the accretion spin-up line on the $P-\dot{P}$ pulsar diagram, and
calculated the spin-down age
 \begin{equation}
 T_{\rm sd}~=~\frac{P}{(n-1)\dot{P}}~\left[1-\left(\frac{P_0}{P}\right)
^{n-1}\right], 
 \end{equation} 
 where $n$ is the braking index. For $n=3$, they derive spin down ages of
65\,Myr and 200\,Myr, for PSR~B1913+16 and PSR~B1534+12, respectively,
lower than their characteristic ages by 40\% and 20\%. These spin-down
ages increase by a factor of 1.7 as n varies from 3 to 2.  Their
sensitivity to the position of the spin-up line on the $P-\dot{P}$ plane
is found to be much weaker. We consider these spin-down ages to be more
reliable as they are better physically motivated. Further, for
PSR~B1913+16, kinematic constraints can be derived based on its position
in the Galaxy and its measured transverse velocity. \cite{W00}(2000)
derived a minimum kinematic age of 35\,Myr, for one intersection with the
Galactic plane after its formation (\cite{A98}1998 derived a higher value
for the minimum kinematic age of 60\,Myr, but this is the result of their
simplifying assumption of a constant Galactocentric radius, which breaks
down for ages longer than just a few Myr).  This minimum kinematic age is
consistent with both the characteristic and spin-down age of PSR~B1913+16.
Taking into account (i) the dependence of the spin-down ages on the
braking index $n$ and on the position of the spin-up line, and (ii) the
kinematic constraints for PSR~B1913+16, we obtain conservative errors on
$T_{\rm sd}$: $65_{-30}^{+45}$\,Myr, and $200_{-40}^{+140}$\,Myr, for
PSR~B1913+16 and PSR~B1534+12, respectively.

The remaining ``observable'' lifetimes of pulsars in coalescing NS--NS
binaries are limited by a number of different factors. It is thought that
pulsars cease to emit pulsed radiation as they cross the empirical ``death
line'' on the $P-\dot{P}$ pulsar diagram ($\dot{P}P^{-5}\simeq
5\times10^{-17}$\,s$^{-5}$; \cite{R76}1976; \cite{MT77}1977;
\cite{LGS98}1998). This line represents a constant magnetic field strength
at the light cylinder and can be theoretically justified in the context of
pulsar emission models in which the emission source is close to the light
cylinder. The time to reach the death line sets a limit to the remaining
observable lifetimes. However, for the recycled pulsars in these systems
this ``death-line'' lifetimes are rather long ($2\times 10^{10}$\,yr and
$6.7\times 10^9$\,yr for PSR~B1534+12 and PSR~B1913+16, respectively) and
do not lead to interesting constraints. 

The merger event itself sets an upper limit to the lifetimes of the
binaries.  Merger times can be calculated accurately, given the measured
binary characteristics and the rate of angular momentum loss due to
gravitational radiation. They are $2.7\times10^9$\,yr and
$3\times10^8$\,yr for PSR~B1534+12 and PSR~B1913+16, respectively and have
been used in the studies of \cite{N91}(1991), \cite{P91}(1991), and
\cite{CL95}(1995). More recently, \cite{vdHL96}(1996) argued that the
lifetime of the radio pulsars in the NS--NS are more strongly limited by
luminosity evolution. They calculated the time at which the magnetic
dipole energy loss rate (pulsar luminosity being proportional to this)
drops below the level reached by a normal, young pulsar in its lifetime
(assumed to be $10^7$\,yr). The implicit assumption in this constraint is
that the pulsar becomes undetectable when it becomes too faint. Although
this is true given the flux limited pulsar surveys, it appears to be
redundant since a correction for the faint end of the pulsar luminosity
function, i.e., pulsars fainter than the minimum flux survey limits, is
also applied to the birth rate estimate later.

\cite{A98}(1998) have pointed out one last constraint imposed on the
observable lifetime. It is related to the known selection effect of pulsar
searches against pulsars in tight binaries. For tight orbits (orbital
periods of a few hours) the Doppler shift of the pulse due to the rapidly
changing acceleration of the pulsar leads to significant signal-to-noise
reduction. The two systems of interest here could not have been detected
with the completed surveys to date if their orbital periods were half
their present values (\cite{A98}1998). PSR~B1913+16 and PSR~B1534+12 will
have half the current orbital periods in $1.8\times10^8$\,yr and
$2.2\times10^9$\,yr, respectively.

We adopt the derived spin--down ages as the current ages of the two
coalescing systems, and the times to reach half the current orbital period
as their remaining lifetimes. These latter are by definition shorter than
the merger times, although the difference between the two for the observed
systems is quite small. The total lifetimes derived for PSR~B1913+16 and
PSR~B1534+12 are $\simeq 2.5\times 10^8$\,yr and $\simeq 2.5\times
10^9$\,yr, respectively. Note that the remaining lifetimes (times to reach
half the current orbital periods) are considerably longer and much more
accurate than the spin--down age estimates, and therefore the errors of
the total lifetimes become rather small (less than 10\%). These longer
remaining lifetimes also indicate that the two binary pulsars are still
relatively young. In fact, this is qualitatively expected based on
considerations of pulsar luminosity evolution: it is easier to detected
them early in their lifetime when they are still relatively bright.
In \S\,3.5 we present an analysis of the necessary corrections for
luminosity evolution and faint pulsars and of the associated
uncertainties.

\subsection{Beaming Fraction}

Detection of pulsars relies primarily on the pulsar beam of radiation
intersecting the observer's line of sight. It has long been recognized
that an {\em upward} correction must be applied to the empirical estimates
of NS--NS coalescence rates to account for the fraction of pulsars that
remain undetected not because of survey selection effects but because they
do not beam in our direction (\cite{N91}1991; \cite{P91}1991). This
correction factor depends on the distribution of pulsar beam sizes which  
may be different for different types of pulsars (e.g., recycled or        
non--recycled). Studies of the pulsar emission geometry of mostly
non--recycled pulsars have lead to the development of a number of spin
period--dependent beaming models (e.g., \cite{NV83}1983; \cite{N87}(1987);
\cite{LM88}1988;  \cite{B90}1990). The derived corrections for beaming
fraction typically lie in the range $1-5$ and major uncertainties remain
primarily because of our limited understanding of the pulsar emission
process and geometry and because of limited data available for millisecond
pulsars.

Previous studies of the NS--NS coalescence rate have not attempted to
address this question in any detail and have either adopted a nominal 
value of $3$ for the beaming correction factor or treated it as an unknown
parameter.

We address this issue based on recent analyses of data on pulse
profile--time evolution (with a long baseline) and polarization for
PSR~B1913+16 (\cite{K98}1998; \cite{T99}1999; \cite{KK00}2000;
\cite{WT00}2000) and PSR~B1534+12 (\cite{A96}1996; \cite{S00}2000).  It
has now been confirmed by several groups that the PSR~B1913+16 pulse
profile is evolving with time at more than one frequency and that this
time evolution can be successfully explained by pulsar precession due to
misaligned spin and orbital angular momentum axes (for the origin and
implications of this misalignment, see \cite{W00}2000). \cite{WT00}(2000)  
analyzed a data set that includes recent high-accuracy data from the
Arecibo observatory and found evidence for an elongated beam along the
latitude direction with an elongation factor of $R=1.62$. The
half--opening angle in the latitude direction is found to be
$\rho=12^{\circ}.4$ and the magnetic inclination angle (relative to the
spin axis) is $\alpha=156^{\circ}$. For PSR~B1534+12 too, pulse--profile
evolution has been confirmed, but the data are limited compared to
PSR~B1913+16 and do not allow a detailed analysis. A simpler analysis
(using the rotating--vector model, see, e.g., \cite{RC69}1969;
\cite{LM88}1988) gives a beam half--opening angle of $\rho=4^{\circ}.87$
and $\alpha=114^{\circ}$ (\cite{A96}1996). Future observations of
PSR~B1534+12 will extend the time baseline and could provide us with a
better estimate of its beaming fraction. 

Taking into account both pulsar beams, the beaming fraction is given by   
 \begin{equation}
 f_b~=~\left[\frac{2}{4\pi}\,\int_{0}^{2\pi}\,d\phi\,\int_{\alpha-\rho}^
{\alpha+\rho}\,\sin\theta\,d\theta\,\right]^{-1}.
 \end{equation}
 We derive beaming fractions of 5.8 and 6.5, for PSR~B1913+16 and
PSR~B1534+12, respectively. Based on these, we adopt an upward correction
factor of 6 for pulsar beaming, and we apply it to the rate estimates we
have obtained so far.

\subsection{Faint Pulsars and Small-Number Observed Sample}

Given that the estimates of the coalescence rate are based on the observed
sample of coalescing NS--NS binaries and that pulsar surveys are
flux-limited, it is necessary to include an upward correction to the rate
estimate for the low-luminosity pulsars that are possibly not represented
in the sample.

\cite{CL95}(1995) applied a simple correction (of a factor of 10)
extrapolating an assumed pulsar luminosity function from the luminosity of
the faintest object in the sample (PSR~B1534+12) down to an assumed
luminosity cut-off of $\sim 1$\,mJy\,kpc$^2$. This correction probably
represents an overestimate of the ``missed'' pulsars, since the
detectability of faint pulsars depends on their flux (distance {\em and}
luminosity), while the correction factor introduced by \cite{CL95}(1995)
implies that {\em all} pulsars fainter than the observed ones are
non-detectable regardless of their position in the Galaxy.

A more accurate correction factor was introduced by
\cite{A98}(1998), who convolved the detection volume integral
with the luminosity function. However, with their method of
obtaining a rate estimate they do not make use of any information
on the observed sample or any weighting based on their Galactic
distribution, and assume an average lifetime for the whole NS--NS
population.

Here we present a statistical calculation of the ``faint--pulsar''
correction factor and its uncertainties. It applies to the basic method of
calculating scale factors for each observed objects and is sensitive to
the number of objects and how well these objects represent the luminosity
function of the total Galactic population.

\subsubsection{Model}

We consider {\em model} pulsar populations (characterized by an assumed
luminosity function) and a large number of {\em model} observed samples.
Using a luminosity-dependent model for pulsar selection effects, we apply
the method of scale factors to the {\em model} observed samples (as it is
done with the real observed sample)  and calculate the total {\em
estimated} number of pulsars in the Galaxy, which is equal to the sum of
the scale factors calculated for each pulsar in the observed sample. We
obtain the correction factor for low-luminosity objects and the associated
uncertainties based on a comparison between the estimated total number of
pulsars and the assumed ``true'' population.

The population of both young radio pulsars and millisecond pulsars appears
to be well described by a power-law luminosity function, $\phi (L)$, with
a negative power--law index (\cite{L85}1985; \cite{CC97}1997):
 \begin{equation}
 \phi (L) \propto L^{-p}.
 \end{equation}
 Assuming a minimum luminosity, \Lmin , for pulsars and that $p>1$ we
normalize \fL \, to the total number of pulsars in the Galaxy, \Ng :
 \begin{equation}
 \phi (L)~=~(p-1)\,L_{\rm min}^{p-1}\,N_G\,L^{-p},\hspace{2.5cm} L\geq
L_{\rm min}, 
 \end{equation}
 where we have assumed that the distribution extends to very large
luminosities. The normalization of \fL \, and in general the statistics of
the population are insensitive to the exact value of the maximum
luminosity. \cite{CC97}(1997) presented a maximum likelyhood analysis of
the properties of millisecond pulsars and obtained best-fit values for $p$
and \Lmin : $p=2\pm 0.2$ and $L_{\rm min}=1.1_{-0.5}^{+0.4}$\,mJy\,kpc$^2$
(for non--recycled pulsars, it is $p\gtrsim 1$; see \cite{L85}1985;
\cite{L93}1993). For our standard case, we use $p=2$ and $L_{\rm
min}=1$\,mJy\,kpc$^2$. In our parameter study we examine the sensitivity
of the results to these values.

The scale factors, ${\cal S}$, as a function of spin period and
luminosity, have been calculated in detailed studies of pulsar populations
and selection effects (e.g., \cite{N87}1987). The dependence on luminosity
appears to be well described by three segments: (i) At high luminosities
pulsars can be detected essentially anywhere in the Galaxy, so \SL $=1$;
(ii) At intermediate luminosities detection is volume-limited in a
disk-like population, so \SL $\propto L^{-1}$; (iii) At sufficiently low
luminosities the limiting distance from the Sun is smaller than the disk
height and detection is volume limited in a homogeneous population, so \SL
$\propto L^{-3/2}$. By matching the segments at the luminosity boundaries
we obtain
 \begin{eqnarray}
 {\cal S}(L) & = & L_1^{1/2}\,L_2\,L^{-3/2} \hspace{1.in} L_{\rm min} <
L
< L_1 \nonumber \\
  & = & L_2\,L^{-1} \hspace{1.75in} L_1 < L < L_2 \nonumber \\
  & = & 1 \hspace{2.15in} L_2 < L
 \end{eqnarray}
 Guided by the results of \cite{N87}(1987) we choose, for our standard
case, $L_1=30$\,mJy\,kpc$^2$ and $L_2=3000$\,mJy\,kpc$^2$.  In what
follows we ignore the dependence of \SL \, on pulsar spin period. This
simplification is probably well justified, since the pulsars found in
NS--NS binaries cover a relatively narrow range of spin periods
($40-60$\,ms)  and the variation of \SL \, with period is rather small
(\cite{N87}1987).

The luminosity function of {\em observed} pulsars, \NL , is then given
by \NL $=$\fL $/$\SL \,and we can calculate the mean number of observed
pulsars as a function of \Lmin , $L_1$, $L_2$ , and $p$, using equations
(2) and (3):
 \begin{eqnarray}
 <N_{\rm obs}>&=&\int_{L_{\rm
min}}^{\infty}\,N(L)\,dL \nonumber \\
 &=& 
N_G\,[(p-1)\,\left(p-\frac{5}{2}\right)^{-1}\,L_2^{-1}\,\left(L_{\rm
min}^{3/2}\,L_1^{-1/2}\,-\,L_{\rm 
min}^{p-1}\,L_2^{2-p}\right) \nonumber \\
 & & +\,\frac{p-1}{p-2}\,L_{\rm
min}^{p-1}\,L_2^{-1}\,\left(L_1^{2-p}\,-\,L_2^{2-p}\right)\,+\,L_{\rm 
min}^{p-1}\,L_2^{1-p}]. 
 \end{eqnarray} 
 The case of $p=2$ is singular for the above equation. For $p=2$, 
$<N_{\rm obs}>$ is given by
 \begin{equation}
 <N_{\rm obs}>~=~N_G\,\left[2\,L_2^{-1}\,\left(L_{\rm min}-L_{\rm
min}^{3/2}\,L_1^{-1/2}\right)~+~L_{\rm min}\,L_2^{-1}\,\ln
(L_2/L_1)~+~L_{\rm min}\,L_2^{-1}\right], 
 \end{equation}
 which for our the choice of \Lmin , $L_1$, $L_2$ in our standard case
becomes: $<N_{\rm obs}>\simeq N_G/415$.

It can be shown that, in this simple (proof-of-principle) model, the mean
estimated total number of pulsars is actually equal to the true total
number of pulsars in the Galaxy. The mean scale factor for the observed
population is
 \begin{equation}
 <{\cal S}(L)>~=~\frac{\int_{L_{\rm min}}^{\infty}\,{\cal
S}(L)\,N(L)\,dL}{\int_{L_{\rm min}}^{\infty}\,N(L)\,dL}~=~
\frac{\int_{L_{\rm min}}^{\infty}\,\phi (L)\,dL}{<N_{\rm obs}>}~=~
\frac{N_G}{<N_{\rm obs}>}.
 \end{equation}
In the case of an observed sample, the mean scale factor can be
expressed in a discrete form: 
 \begin{equation} 
 <{\cal S}(L)>~=~\frac{\sum _i {\cal S}_i(L_i)}{N_{\rm
obs}}~=~\frac{N_{\rm est}}{N_{\rm obs}}. 
 \end{equation}
From equations (18) and (20), we obtain $N_{\rm est}~=~\frac{N_G}{<N_{\rm
obs}>}\,N_{\rm obs}$ and its mean value then is $<N_{\rm est}>~=~N_G$. We
confirm this equality in our numerical results described in the next
section (see also equation [A8] in the Appendix).

\subsubsection{Calculations and Results}

We use the above model for the pulsar properties and selection
effects and perform Monte Carlo simulations to investigate the
``faint--pulsar'' correction to the rate estimate and its
dependence on the number of objects in the observed sample.

For a given value of the mean number of observed pulsars,
$<N_{\rm obs}>$, we take $N_{\rm obs}$ to be a random number
drawn from a Poisson distribution with a mean equal to $<N_{\rm
obs}>$. We generate a large number of model observed samples consisting of
$N_{\rm obs}$ objects. For each object in the sample, we choose a
luminosity $L_i$ drawn from the observed luminosity function, \NL . We
then evaluate the individual scale factors, ${\cal S}_i(L_i)$, and the
{\em estimated} total number of pulsars in the Galaxy:
 \begin{equation}
 N_{\rm est}~=~\sum_{i=1}^{N_{\rm obs}}\,{\cal S}_i(L_i).
 \end{equation}
 For a given value of $<N_{\rm obs}>$, we repeat the calculation
for a large number ($10^5$) of values of $N_{\rm obs}$, and we
obtain the distribution of $N_{\rm est}$ as a function of
$<N_{\rm obs}>$. To characterize this distribution we use its
mean, median, first quartile (25\%) and third quartile (75\%).

The {\em model} estimated total number $N_{\rm est}$ (eq.\ 20) is
obtained in the same way as it is done for the real observed
sample. The difference is that, in the model calculations, we can
actually compare the distribution of $N_{\rm est}$ to the
``true'' total number of pulsars $N_G$ as a function of $<N_{\rm
obs}>$. Our results for our standard case ($p=2$, $L_{\rm
min}=1$\,mJy\,kpc$^2$, $L_1=30$\,mJy\,kpc$^2$, and
$L_2=3000$\,mJy\,kpc$^2$) are shown in Figure 4. As shown
in \S\,3.5.1, the mean of the distribution is equal to the
``true'' total number of pulsars in the Galaxy. For large values
of $<N_{\rm obs}>$ the distribution of $N_{\rm est}$ is narrow
and its median follows closely its mean value, an indication that
the method of estimated the total pulsar number using the scale
factors leads to reliable results. On the contrary, for small
values of $<N_{\rm obs}>$ ($\lesssim 10$), the distribution
becomes wider and the median deviates systematically to values
lower than the total Galactic number of pulsars (the distribution
becomes highly skewed with an extended tail to large values). For
very low $<N_{\rm obs}>$ ($\lesssim 3$), the mean is dominated by
a small number of realizations for the observed sample, since in
more than 75\% of the cases $N_{\rm est}$ lies below the mean.
For $<N_{\rm obs}>=2$, the derived $N_{\rm est}$ could be an
underestimate of the true number by up to 2 orders of magnitude. 

The $N_{\rm est}$--distributions are shown in Figure 5 (in $N_{\rm
est}\,f(N_{\rm est})$ form), for three different choices of $<N_{\rm
obs}>=2,10,50$. The high degree of skewness towards low values of $N_{\rm
est}$ for a small--number sample is evident. The distributions becomes
much more narrowly concentrated around their almost equal mean and median
values for large--number samples. In the appendix we derive analytic
expressions for the moments of the distribution along with its relative
variance and skewness.

The reason for the extreme skewness of the probability distribution of the
estimated number of pulsars is primarily connected to the properties of
their luminosity function. The true population is dominated by faint
pulsars that are hard to detect in pulsar searches. When only a small
number of pulsars is drawn out of such a population, it is more probable
that they will be bright pulsars. Therefore a small--number sample is more
likely dominated by bright pulsars. This leads to large values of the
estimated volume out to which they could be detected, consequently to low
values of their scale factors, and hence their sum, $N_{\rm est}$,
represents an {\em underestimate} of the true total number. 

It becomes evident that a {\em upward} correction factor must be applied
to the estimated rate to account for this effect. We calculate the
correction based on the ratio of the ``true'' total number of objects to
the estimated total number, in our model. The reasons for this correction
factor are actually equivalent to those of the corrections for
low--luminosity pulsars, adopted by authors in previous studies: for
example, an upward correction of a factor of 10 was introduced by
\cite{CL95}(1995) based on a simple extrapolation of the luminosity
function down to a minimum luminosity. The analysis we present here shows
that this extension to low luminosities is actually quite uncertain (for
reasons other than just the uncertainties in the values of $p$, $L_{\rm
min}$, $L_1$, and $L_2$). The uncertainty rapidly increases for
small--number samples and can be as large as two orders of magnitude for
an observed sample of only two objects.

The above results are quite robust and general as long as the luminosity
function is dominated by faint pulsars and the detection process is such
that faint objects are harder to detect, i.e., flux-limited surveys. The
quantitative details of the degree of underestimation of the true total
pulsar number turn out to be somewhat sensitive to the values of the input
parameters, $p$, $L_{\rm min}$, $L_1$, and $L_2$. We have run simulations
for a set of different values for all these parameters ($p$: 1.8, 2;
$L_{\rm min}$: 1, 3; $L_1$: 10, 30, 100; $L_2$: 1000, 3000) guided by the
results of \cite{CC97}(1997) and \cite{N87}(1987).  From our parameter
study, we have eliminated cases (sets of $p$, $L_{\rm min}$, $L_1$, and
$L_2$ values) in which the normalization between $N_G$ and $<N_{\rm obs}>$
deviates from current expectations, i.e., $N_G/<N_{\rm obs}>\sim 100$, by
a factor larger than $\sim 5$. This constraint originates in the
(order--of--magnitude) estimates of the total number of pulsars in the
galaxy ($\sim 10^5$) and the number of observed pulsars ($\sim 10^3$). For
all the cases that satisfy this constraint, our results for the median,
first, and third quartile of the $N_{\rm est}$--distributions are given in
Table 1 (see also Figure 6). We find that, for all models, a minimum
number of $5-10$ observed objects (7 for our standard case) is necessary
to obtain a median that underestimates the true number by a factor smaller
than 3.  At $<N_{\rm obs}>=2$ the true total pulsar number is
underestimated typically by factors of $\lesssim 2$ up to $\simeq 300$
($\simeq 2-200$ for our standard case with the median at $\simeq 15$).

\subsubsection{Effect of Distance Errors} 

So far we have neglected any uncertainties in the distance
estimates of the pulsars. Pulsar distances are inferred based on
their dispersion measure and an electron-density model (e.g.,
\cite{TC93}1993) and are known to be uncertain. Here we examine
their effect on the ``faint--pulsar'' correction factors derived
in \S\,3.5.2.

We modify the Monte Carlo simulations described above to include
distance errors. For each object in the {\em model} samples, we
generate not only a luminosity $L$ but also a
position in the Galaxy described by a galactocentric radius $R$, a
vertical distance from the Galactic plane $Z$, and an azimuthal
angle $\Phi$. These three parameters are assigned to the observed
objects according to the probability distribution $F_p$
 \begin{equation} 
F_p(R,Z,\Phi)~\propto~\frac{1}{2\pi}\,\exp\left(-\frac{R^2}{2R_0^2}-
\frac{\vert z\vert}{z_0}\right),
 \end{equation} 
 where $R_0=4$\,kpc and $z_0=1$\,kpc are the radial and vertical scale
lengths. Given the position of the pulsar, we calculate its ``true''
distance $D$ from the Sun. We then generate in our Monte Carlo simulations
an ``inferred'' distance $D'$.  This distance is drawn from a Gaussian
distribution with a mean equal to $D$ and a standard deviation
$\sigma=fD$, where $f$ is a factor smaller than unity: $f(D')\propto
\exp[-(D'-D)^2/(2(fD)^2)]$. We use the two distances and the true
luminosity and assign an ``observed'' luminosity $L'=L\,(D'/D)^2$ to each
of the Monte Carlo ``observed'' pulsars. Scale factors (eq.\ [16]) are
calculated using $L'$ and from them we obtain $N_{\rm est}$ (eq.\ [20]).

We obtain results on the distribution of $N_{\rm est}$ compared to the
model ``true'' total number of pulsars in the Galaxy for the case of 20\%
distance errors ($f=0.2$) and the extreme case of 80\% errors
($f=0.8$)\footnote{We have explored the dependence of our results on $R_0$
and $z_0$ by running models for $R_0=8$\,kpc and $z_0=3$\,kpc and the
changes are negligible.}.  We find that errors at the 20\% level are too
small ($\lesssim 1$\%) to have any effect on the $N_{\rm
est}$--distributions. The effect starts becoming non-negligible at
$f\gtrsim0.5$ and is rather significant (factor of 2--3)  for $f=0.8$. For
such large $f$-values, the distribution of $N_{\rm est}$ becomes broader
leading ultimately to highly uncertain empirical rate estimates.  
However, we do not expect that such large distance errors are realistic,
for the general pulsar population. Distance estimates based on the
\cite{TC93}(1993) electron density model have been found to have typical
errors of 20--30\% and therefore we conclude the ``faint--pulsar''
correction factors should not be increased, because of pulsar distance
errors.

\subsection{Conclusions on the Galactic Coalescence Rate} 

We use the derived scale factors and lifetimes for the two coalescing
systems, to obtain an estimate of the Galactic coalescence rate, which we
can further correct for beaming and the faint end of the pulsar luminosity
function. We also identify the most dominant sources of the associated
uncertainty.

Based on our results on the NS--NS scale height we obtained ${\cal
S}_{1913+16}=45-60$ and ${\cal S}_{1534+12}=145-200$.  We combine these
results with the estimated lifetimes of the two systems and obtain an
estimate for the NS--NS coalescence rate in the range $2-4\times
10^{-7}$\,yr$^{-1}$.

This Galactic rate must be further corrected for the fraction of
coalescing NS--NS binaries with pulsars that do not beam in our direction.
We found this {\em upward} correction factor to be $\simeq 6$ leading to
rates in the range $(1-2.5)\times 10^{-6}$\,yr$^{-1}$. We further have to
correct the rate for the fraction of pulsars at the faint end of the
luminosity function that are hard to detect and are not represented in the
observed sample. For low-number observed samples ($\lesssim 5-10$), in
particular, this ``faint-pulsar'' correction factor has a broad
distribution and typically contributes to an {\em upward} revision of the
rate. The correction factors are found to be as high as $\sim 200$ (see
Table 1). As a result we estimate the Galactic NS--NS coalescence rate to
lie in the range $\simeq 10^{-6} - 5\times 10^{-4}$\,yr$^{-1}$, where the
extent of this range is primarily dominated by the uncertainty in the
``faint--pulsar'' correction for a small--number sample.  

In our analysis we have obtained estimates and the associated
uncertainties for each of the relevant factors, separately. Therefore, it
will be straightforward to update our coalescence rate estimate, if new
and improved estimates of ${\cal S}_{1913+16}$ and ${\cal S}_{1534+12}$
are obtained in the future.

\section{EXTRAGALACTIC RATE EXTRAPOLATION}

Interferometric ground--based gravitational--wave observatories currently
under construction are expected to be sensitive to NS--NS inspiral events
out to extragalactic distances. Specifically for LIGO, the most
up--to--date estimates of the maximum detection distances place them at
$\simeq 20$\,Mpc and $\simeq 350$\,Mpc, for LIGO I and II, respectively
(\cite{G99}1999; Finn 2000, private communication).  Hence, we need to
extrapolate the Galactic coalescence rate estimated in \S\,3 out to the
volume of the Universe accessible to LIGO I and II.  There are several
ways of performing this extrapolation, using, for example, scalings with
galaxy mass, luminosity, or star formation rate. The implicit assumption
in such types of scalings is that (1) the NS--NS inspiral rate is
proportional to the formation rate of massive stars, (2) the properties
(mass function, binary fraction, etc.) of the primordial binary population
in galaxies (within distances of interest) are not grossly different than
those of the Galactic population, and (3) the star formation history out
to these distances has been roughly uniform, i.e., the fractions of
starburst galaxies and ellipticals are rather small.

\cite{P91}(1991) extrapolated the Galactic inspiral rate based on
estimates of the B--band luminosity density of the universe obtained from
galaxy counts (\cite{E88}1988) and the B--band luminosity of the Milky Way
($\simeq 2\times 10^{10}$\,L$_{\odot,B}$). He also included an upward
correction, for any reprocessed B--band luminosity that is emitted in the
infrared, and a downward correction, for the contribution of E,S0 galaxies
to the B--band luminosity, since E,S0 galaxies stopped forming stars
billions of years ago and are not likely to produce very many inspiral
events at present. \cite{N91}(1991) obtained a different extrapolated
estimate based on the average galaxy number density out to a few hundred
Mpc, whereas \cite{CL95}(1995) used the number density of galaxy clusters,
number of galaxies per cluster, and average size of voids between
superclusters. The last two approaches include the additional assumption
that each galaxy out to the distances of interest has the same NS--NS
inspiral rate as the Milky Way.

In the present study we adopt a method of extrapolation similar to that
used by \cite{P91}(1991), but with up--to--date estimates of the B--band
and infrared (IR) luminosity densities of the nearby universe and of the
Milky Way B--band luminosity.

Luminosity functions have been derived for the most recent, completed, and
relatively large--volume galaxy surveys both in the blue, the Stromlo--APM
Redshift Survey with a sample median redshift of 0.05 (\cite{L92}1992),
and in the red, the Las Campanas Redshift Survey with a sample median
redshift of 0.1 (\cite{L96}1996). Although the details of the two
luminosity functions (after corrections for the difference in colors) are
different, the integrated luminosity densities are found to be very
similar, equal to $\simeq (1.4\pm 0.1)\times
10^8\,h$\,L$_\odot$\,Mpc$^{-3}$ ($h\equiv
H_0/100$\,km\,s$^{-1}$\,Mpc$^{-1}$). This result is also consistent with
preliminary results from the Sloan Digital Sky Survey (\cite{B00}2000). In
contrast to \cite{P91}(1991), we do not include a downward correction for
the fraction of E,S0 galaxies because it is balanced by a similar
correction of the Galactic B-band luminosity for the contribution from the
bulge. However, following \cite{P91}(1991), we do apply a correction for
the fraction of the infrared luminosity density due to re-processed blue
light by dust. We use the far--infrared ($40-100\mu$m)  luminosity density
derived by \cite{S90}(1990) for a sample of IRAS galaxies, and we further
increase it by 30\% to account for emission down to $12\mu$m (see
\cite{SN91}1991): $L_{\rm IR}\simeq (3.6\pm 0.4)\times
10^7\,h$\,L$_\odot$\,Mpc$^{-3}$. We also adopt results of spectral
syntheses (following \cite{P91}1991) showing that in the absence of any
extinction half of this infrared light comes from reprocessed blue light.
Then, depending on the assumed extinction law, we obtain an ``intrinsic''
B-band luminosity density in the range: $1.5-2\times
10^8\,h$\,L$_\odot$\,Mpc$^{-3}$. Note that the statistical uncertainties
of the luminosity densities are overshadowed by the uncertainties in the
dust extinction law.

For the B--band luminosity of our Galaxy, we obtain an estimate using the
B--band Tully--Fisher relation from \cite{Y97}(1997) (their equation [5])
and adopting a circular velocity at the solar radius equal to
220\,km\,s$^{-1}$ (\cite{FW97}1997). This circular velocity is also
consistent with the Galactic potential used in \S\,3.2, which gives a
velocity of 222\,km\,s$^{-1}$ at 8\,kpc from the Galactic center). We
calculate the blue absolute magnitude of the Milky Way to be -20.11, which
implies that the Galactic B--band luminosity equal to $9\times
10^9$\,L$_\odot$. This value is also consistent with that of NGC 891, a
galaxy very similar to the Milky Way with the same circular velocity. Our
estimate of the blue luminosity of the Galaxy is about a factor of 2 lower
than the one used by \cite{P91}(1991). His adopted value is taken from a
study by \cite{vdK87}(1987), where a rather large value of the disk radial
scale length was assumed (5\,kpc). Studies of the Galactic disk in the IR
lead to scale length estimates in the range $2.3-2.6$\,kpc
(\cite{F98}1998; \cite{DS00}2000), whereas corresponding estimates in the
blue lead to values higher by about 30\% {\cite{dJ94}1994). These low
values are also consistent with studies of disk kinematics (see
\cite{DB98}1998). Since the total luminosity scales roughly with the
square of the scale length, these lower estimates are consistent with our
lower (by a factor of 2) estimate of the blue luminosity of the Milky Way.

Based on the above estimates and corrections, we derive the scaling factor
for the extragalactic extrapolation of the Galactic rate, i.e., the ratio
of the B-band luminosity density to the B-band luminosity of the Milky Way
in the range: $(1-1.5)\times 10^{-2}$\,Mpc$^{-3}$, for $h=0.65$ (the
effective maximum distances of reach for LIGO I and II have been
calculated for $h=0.65$).

In principle, another way of obtaining an extrapolation factor is to use a
scaling based on star formation rate estimates. However, at present these
are much more uncertain than the results obtained from luminosity
densities. For comparison, we derive a scaling factor using (i) a Galactic
star formation rate of $1-3$\,M$_\odot$\,yr$^{-1}$, which includes
recycling of molecular gas (\cite{B97}1997; \cite{LF85}1985), and (ii)  a
star formation density in the local Universe of $1.2-4\times
10^{-2}$\,M$_\odot$\,yr$^{-1}$\,Mpc$^{-3}$ (for $h=0.65$; \cite{M98}1998;
\cite{ST00}2000). Consequently, we obtain a highly uncertain scaling
factor to lie in the range $0.4-4\times 10^{-2}$\,Mpc$^{-3}$.

In what follows, we adopt the scaling for the extragalactic extrapolation
based on the B-band luminosities. 

\section{DISCUSSION}

We use the observed sample of close NS--NS binaries that will coalesce in
less than $10^{10}$\,yr to estimate the Galactic NS--NS coalescence. We
study in detail the uncertainties involved in this estimate and we find
that the coalescence rate lies in the range: $10^{-6}-5\times
10^{-4}$\,yr$^{-1}$. We also calculate a simple scaling law for the
extrapolation of this rate to extragalactic distances, based on the
B--band luminosities of galaxies. These combined with the maximum
detection distances that are expected to be reachable with LIGO (20\,Mpc
and 350\,Mpc for LIGO I and II, respectively, calculated for $h=0.65$;
Finn 2000, private communication), lead us to estimates of the NS--NS
inspiral detection rates, for LIGO I and II, in the ranges $3\times
10^{-4} - 2.5\times 10^{-1}$\,yr$^{-1}$ and $2-1300$\,yr$^{-1}$,
respectively. We conclude that the prospects for detection of such
inspiral events by LIGO II are strongly encouraging.

In this study we have given special attention to the various uncertainties
associated with these empirical rate estimates and tried to quantify them
wherever possible. We find that, by a very large margin, the most dominant
source of uncertainty is the necessary correction for undetectable pulsars
at the faint end of the luminosity function (see \S\,3.5). Because of the
small number of objects in the observed sample, this correction factor
covers a wide range of values and can be as high as $\sim 200$.
Uncertainties originating from the dynamical evolution of the radio pulsar
population and pulsar lifetimes are within a factor of 2, whereas
uncertainties in the beaming fraction and the extragalactic
extrapolation are found to be less important ($\lesssim 50\%$).

It is evident that any expectations for reducing this
two--orders--of--magnitude uncertainty in the empirical coalescence rate
strongly depend on the discovery of more close NS--NS binaries. We have
shown that if the sample increases to $\sim 10$ objects then the
correction for the faint end of the radio luminosity function can be
reduced to just a factor of a few or less (see Figures 4 \& 6). The
ongoing Parkes Multibeam Survey (\cite{L00}2000) could possibly contribute
to the increase of the sample of coalescing NS--NS systems. PSR~J1141-6545
was initially considered as a candidate NS--NS but follow--up measurements
of relativistic parameters imply a total mass for the binary system low
enough that the hypothesis of a NS--WD binary instead of a NS--NS binary
appears to be more probable (\cite{Kas00}2000). We note that, based on the
characteristic age of this pulsar (1.4\,Myr) and the time to reach the
pulsar ``death--line'' (29\,Myr), were this system a NS--NS binary, its
contribution to the coalescence rate would be ${\cal S}\times 3.3\times
10^{-8}$\,yr$^{-1}$, where ${\cal S}$ would be its scale factor.  To our
knowledge, only about half of the complete Parkes Multibeam Survey data
set has been analyzed (although more extensive acceleration searches will
follow), so it is reasonable to expect that new NS--NS binaries could be
discovered in the next few years.

We note that the rate estimates we derive here are relevant to NS--NS
binaries with recycled, hence long--lived radio pulsars, like the ones
detected in the observed systems. Recent results of theoretical
calculations of NS--NS formation indicate that systems can form without
any of the two NS having a chance to be recycled by accretion
(\cite{BK01}2001), and that a significant population of non-recycled
NS--NS binaries with very short radio--pulsar lifetimes can exist in the
Galaxy. The possible existence of such a population implies an increase of
the NS--NS coalescence rate by small but non-zero factors according to the
results in \cite{BK01}(2001).

Apart from actual estimates of the coalescence rates, observed samples of
radio pulsars have also been used to derive upper limits to the Galactic
coalescence rate. Based on the absence of young radio pulsars found in
NS--NS binaries, \cite{B96}(1996) derived an upper limit of $\sim
10^{-5}$\,yr$^{-1}$. \cite{A98}(1998) raised this limit to $\sim
10^{-4}$\,yr$^{-1}$ using a more detailed analysis of the same basic
argument. These limits were calculated prior to the study by
\cite{BK01}(2001) and they are expected to increase in view of them.
\cite{KL00}(2000) derived a similar upper limit ($\sim
10^{-4}$\,yr$^{-1}$), although based on completely different
considerations, related to the formation branching ratios of radio pulsars
like those in close NS--NS binaries and their possible single
counterparts. It is quite encouraging to see that both limits are in
agreement with each other and the range of coalescence rates we obtained
here is also consistent (within a factor of 5) with these upper limits. If
we adopt these upper limits strictly then the LIGO II detection rate is
reduced to about $2-300$ events per year.

We can also compare our results to the theoretical estimates for the
NS--NS coalescence rates obtained from studies of the evolution of binary
populations until the formation of close NS--NS systems (e.g.,
\cite{L97}1997; \cite{FBB98}1998; \cite{PZ98}1998; \cite{BB98}1998;
\cite{FWH99}1999; \cite{B99}1999; \cite{G01}2001). The results of these
studies have been summarized elsewhere (e.g., \cite{K00a}2000a;
\cite{KL00}2000) and the rates have been found to cover a wide range of
3--4 orders of magnitude. The dominant uncertainties in these studies are
related to multiple supernovae and NS kicks, the treatment of dynamically
unstable mass--transfer episodes, and the properties of the primordial
binary populations (e.g., mass ratios). The accuracy of these purely
theoretical estimates could be improved, if a set of observational
constraints (e.g., different supernova rates, statistics of Wolf--Rayet
populations) are imposed on the models and their absolute normalization
(see \cite{BK00}2000).  It is evident that, at present, studies of the
observed sample of NS--NS binary pulsars provide us with a more accurate
estimate of the NS--NS coalescence rate.

 \acknowledgements
 VK would like to thank T.\ Abel, T.\ di Matteo, C.\ Kochanek, M.\ Kramer,
D.\ Lorimer, C.\ Metzler, and I.\ Stairs for helpful and stimulating
discussions. This work was supported in part by the Smithsonian
Astrophysical Observatory in the form of a Harvard-Smithsonian Center for
Astrophysics Postdoctoral Fellowship and a Clay Fellowship to VK, and in
part by NSF grants PHY 9507695 to RN and AST 96-18357 to JHT.

\appendix

\section{Moments of the Distribution of Estimated Pulsar Numbers}

The estimated number of pulsars $N_{\rm est}$\footnote{In what follows we
use the notation from \S\,3.5} (see eq.\ [22]) can be written as
 \begin{equation}
 N_{\rm est}~=~\sum_i\,N_i~=~\sum_i\,n_i\,s_i, 
 \end{equation}
 where $<n_i>=\frac{\phi(L_i)}{{\cal S}(L_i)}\,\Delta L$ and $\Delta L$ is
the width of luminosity bins. We denote the probability distribution of
$N_{\rm est}$ with $p(N_{\rm est})$ and that of $N_i$ with $p_i(N_i)$. If
$\chi(t)$ and $\chi_i(t)$ are the Fourier transforms of $p(N_{\rm est})$
and $p_i(N_i)$, respectively, then the Fourier convolution theorem implies
that
 \begin{equation}
 \chi(t)~=~\prod_i\chi_i(t). 
 \end{equation}
 We use the characteristic function of a Poisson process for $N_i$ and
therefore: 
 \begin{equation}
 \chi_i(t)~=~\exp\left\{\frac{\phi(L_i)}{{\cal S}(L_i)}\,\Delta
L\left[\exp(i\,{\cal S}(L_i)\,t)-1\right]\right\}. 
 \end{equation}
 Equations (A2) and (A1) imply that
 \begin{eqnarray}
 \log \chi(t) & = & \sum_i \log \chi_i(t) \\
 & = & \sum_i \frac{\phi(L_i)}{{\cal S}(L_i)}\,\Delta 
L\left[\exp(i\,{\cal S}(L_i)\,t)-1\right] \\
 & = & \int dL\,\frac{\phi(L_i)}{{\cal S}(L_i)}\,\left[\exp(i\,{\cal
S}(L_i)\,t)-1\right]
 \end{eqnarray}

 We can then use the moment equation 
 \begin{equation}
 \mu^r~=~\frac{d^r\chi}{d(it)^r \vert _{t=0}}
 \end{equation}
 to compute then mean, root--mean--square, and skewness of the
distribution
 \begin{equation}
 \langle N_{\rm est} \rangle~=~\int dL\,\phi(L)\,=\,N_G, 
 \end{equation}
 \begin{equation}
 \langle N_{\rm est}^2 \rangle~=~\langle N_{\rm est} \rangle^2\,+\,\int
dL\,\phi(L)\,{\cal S}(L),
 \end{equation}
 \begin{eqnarray}
 \langle N_{\rm est}^3 \rangle & = & \langle N_{\rm est}
\rangle^3\,+\,3\,\langle N_{\rm est} \rangle\,\int
dL\,\phi(L)\,{\cal S}(L)\,+\,\int dL\,\phi(L)\,s^2(L) \nonumber \\
  & = & 3\,\langle N_{\rm est} \rangle\,\langle N_{\rm est}^2
\rangle\,-\,2\,\langle N_{\rm est} \rangle^3\,+\,\int dL\,\phi(L)\,s^2(L). 
 \end{eqnarray}
 The above expression can be used to calculate the magnitude of dispersion
of $N_{\rm est}$ relative to its expected value and the skewness of the
$N_{\rm est}$--distribution relative to its dispersion. 

Ultimately we are interested in the magnitude of the variance
and the
skewness and their dependence 
on $<N_{\rm obs}>$. As it can be seen from its definition
in \S\,3.5, ${\cal S}(L)$ does not depend on $<N_{\rm obs}>$, but instead
depends only on $p$, $L_{\rm min}$, $L_1$, and $L_2$ (and $L$ of course).
On the other hand,
$\phi(L)$ is proportional to $N_G$, and hence proportional to $<N_{\rm
obs}>$ (see eqs.\ [16], [18]). Also the integrals in eqs.\
(A8)--(A10) are independent of $<N_{\rm obs}>$. A measure of the relative
dispersion is given by
 \begin{equation}
 \frac{\left[\int\,\left(N_{\rm est}-\langle\,N_{\rm
est}\rangle\right)^2\,f(N_{\rm est})\,dN_{\rm est}\right]^{1/2}}
{\langle\,N_{\rm est}\rangle},  
 \end{equation}
 which is equal to
 \begin{equation}
 \frac{\left[\langle N_{\rm est}^2 \rangle-\langle N_{\rm est}
\rangle^2\right]^{1/2}}{\langle N_{\rm est}\rangle}~=~\frac{\left[\int
dL\,\phi(L)\,{\cal S}(L)\right]^{1/2}}{\int
dL\,\phi(L)}~\propto~\langle\,N_{\rm
obs}\,\rangle^{-1/2}. 
 \end{equation}
 A measure of the skewness of the $N_{\rm est}$--distribution relative to
its dispersion is given by
 \begin{equation}
 \frac{\int\,\left(N_{\rm est}-\langle\,N_{\rm
est}\rangle\right)^3\,f(N_{\rm est})\,dN_{\rm est}}{\left[\langle
N_{\rm est}^2 \rangle-\langle N_{\rm est} \rangle^2\right]^{3/2}}, 
 \end{equation}
 which is equal to
 \begin{equation}
 \frac{\langle N_{\rm est}^3 \rangle\,-\,3\,\langle N_{\rm est}
\rangle\,\langle N_{\rm est}^2 \rangle\,+\,2\,\langle N_{\rm est}
\rangle^3}{\left[\int dL\,\phi(L)\,{\cal S}(L)\right]^{3/2}}~=~\frac{\int
dL\,\phi(L)\,s^2(L)}{\left[\int
dL\,\phi(L)\,{\cal S}(L)\right]^{3/2}}~\propto~\langle\,N_{\rm
obs}\,\rangle^{-1/2}.
 \end{equation}
 It is evident that the relative variance and skewness of the
distribution decrease as the number of systems in the observed sample
increase. 

The proportionality constants in equations (A12) and (A14) provide us with
information about the magnitude of the variance relative to the mean and
the skewness relative to the variance. These constants can be calculated
analytically using equations (16) -- (18). As expected based on our
discussion in \S\,3.5, we find that both the variance and the skewness are
greatly dominated by the faint--pulsar branch of ${\cal S}(L)$ ($L_{\rm
min} < L < L_1$). The relevant analytic expressions are given for the
variance: 
 \begin{eqnarray}
 \frac{\left[\langle N_{\rm est}^2 \rangle-\langle N_{\rm est}
\rangle^2\right]^{1/2}}{\langle N_{\rm
est}\rangle}&=&(p-1)^{1/2}\,\left(p+\frac{1}{2}\right)^{-1/2}\,L_{\rm
min}^{(p-1)/2}\,L_2^{1/2} \nonumber \\
 & &\times \left[L_{\rm
min}^{-(p+1/2)}\,L_1^{1/2}\,-\,L_1^{-p}\right]^{1/2}\,N_G^{-1/2},
 \end{eqnarray}
and the skewness: 
 \begin{eqnarray}
 \frac{\langle N_{\rm est}^3 \rangle\,-\,3\,\langle N_{\rm est}
\rangle\,\langle N_{\rm est}^2 \rangle\,+\,2\,\langle N_{\rm est}
\rangle^3}{\left[\int dL\,\phi(L)\,{\cal S}(L)\right]^{3/2}}&=&
(p-1)^{-1/2}\,(p+2)^{-1}\,\left(p+\frac{1}{2}\right)^{3/2} \nonumber \\
 & & \times L_{\rm min}^{(1-p)/2}\,L_1\,L_2^{1/2}\,
\left[L_{\rm
min}^{-(p+2)}\,-\,L_1^{-(p+2)}\right] \nonumber \\
 & & \times  \left[L_{\rm
min}^{-(p+1/2)}\,L_1^{1/2}\,-\,L_1^{-p}\right]^{-3/2}\,N_G^{-1/2},
 \end{eqnarray}
 where $N_G$ is given by equation (18) as a function of $p$, $L_{\rm
min}$, $L_1$, $L_2$, and is proportional to $<N_{\rm obs}>$.

For our standard case (model \#\,1 in Table 1), we obtain from equations
(A15), (A16), and (19):
 \begin{displaymath}
 \frac{\left[\langle N_{\rm est}^2 \rangle-\langle N_{\rm est}
\rangle^2\right]^{1/2}}{\langle N_{\rm
est}\rangle}~\simeq~4\,\langle\,N_{\rm
obs}\,\rangle^{-1/2}, 
 \end{displaymath}
 and 
 \begin{displaymath}
 \frac{\langle N_{\rm est}^3 \rangle\,-\,3\,\langle N_{\rm est}
\rangle\,\langle N_{\rm est}^2 \rangle\,+\,2\,\langle N_{\rm est}
\rangle^3}{\left[\int dL\,\phi(L)\,{\cal
S}(L)\right]^{3/2}}~\simeq~6\,\langle\,N_{\rm
obs}\,\rangle^{-1/2}. 
 \end{displaymath} 
 Using our Monte Carlo simulations we were able to confirm our analytical
results for both the variance and the skewness to an accuracy better than
1\%. A comparison of numerical and analytical results is shown in Figure
7, for our standard case. We note that the above analytic expressions can
be used to update the results on the variance and the skewness of the
$N_{\rm est}$--distribution as our knowledge of the luminosity
function improves.

\newpage

 \centerline{\psfig{figure=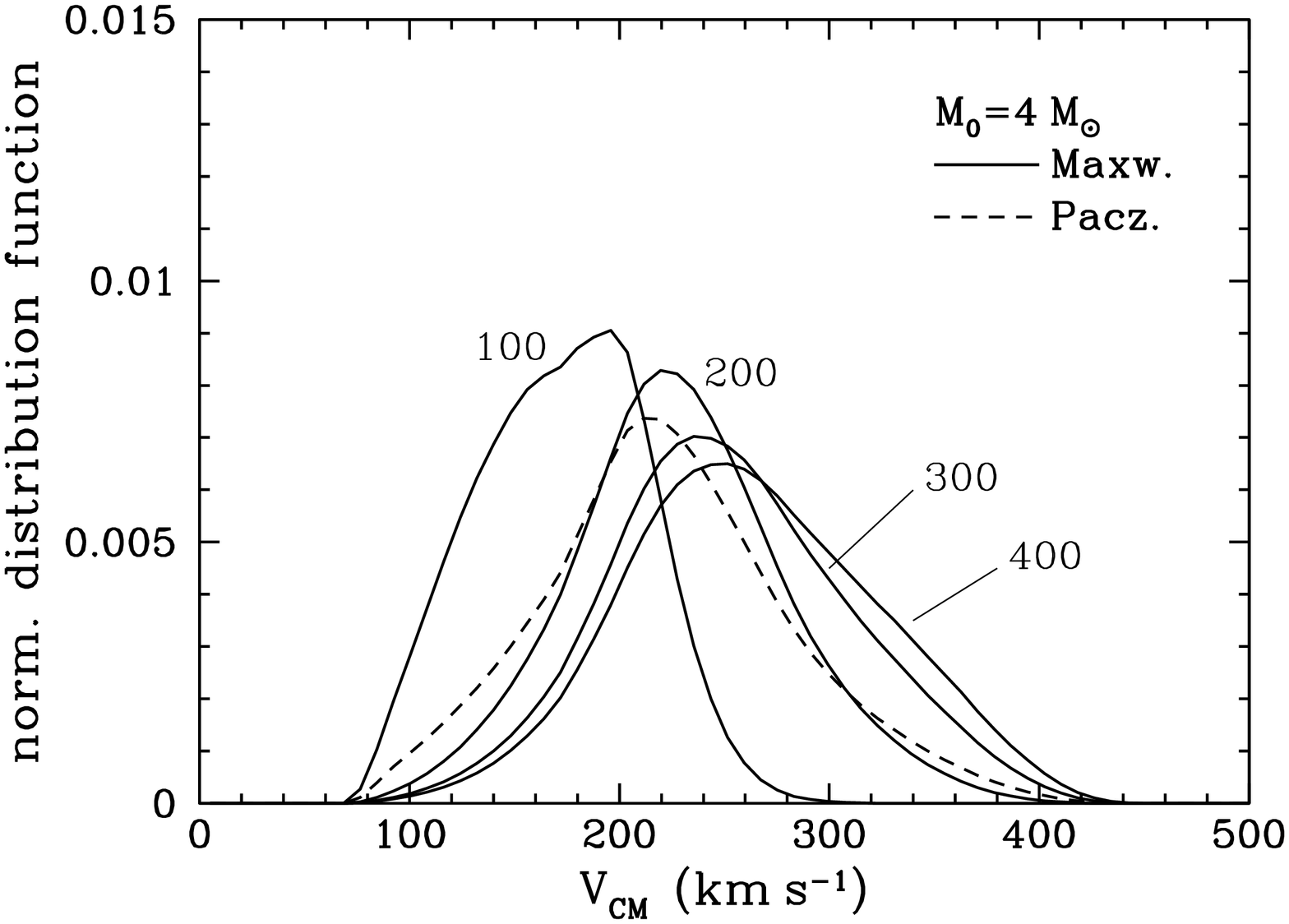,angle=-90,width=6.5in}}
 \figcaption{Probability distributions $f(V_{\rm CM})dV_{\rm CM}$ of
center--of--mass velocities $V_{\rm CM}$ of coalescing NS--NS binaries
just after the second supernova explosion, shown for a specific
helium--star mass ($M_0=4$\,M$_\odot$). Solid lines correspond to models
with a Maxwellian kick--magnitude distribution with
$\sigma=100,200,300,400$\,km\,s$^{-1}$, and the dashed line corresponds to
a ``Paczynski--like'' kick--magnitude distribution (see text). NS kicks
are assumed to be isotropic.}

\centerline{\psfig{figure=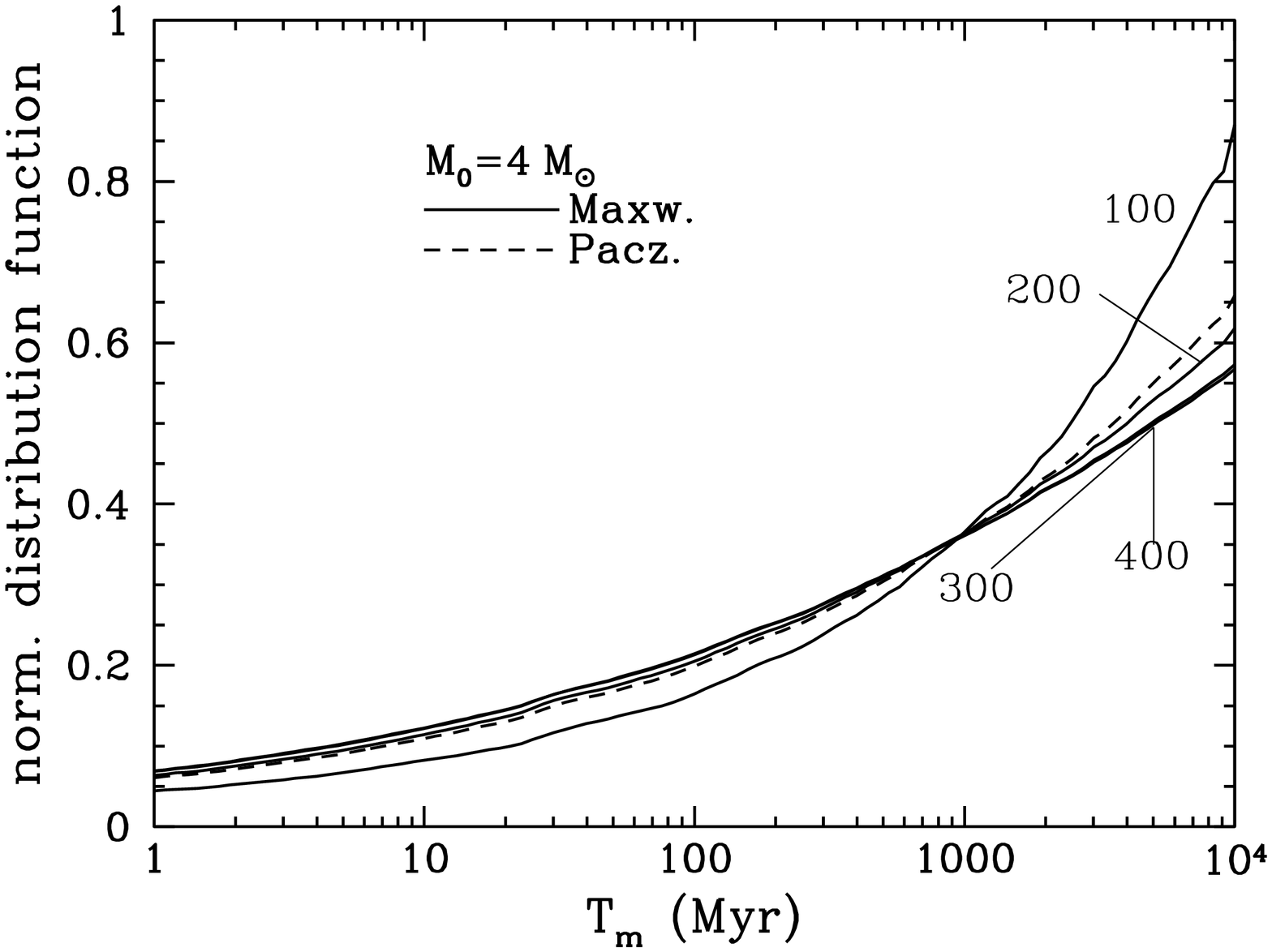,angle=-90,width=6.5in}}
 \figcaption{Probability distributions $f(\log T_m)d\log T_m)$ of merger
time scales $T_m$ of coalescing NS--NS binaries, for a specific
helium--star mass ($M_0=4$\,M$_\odot$). Line coding is as in Figure 1.}

\centerline{\psfig{figure=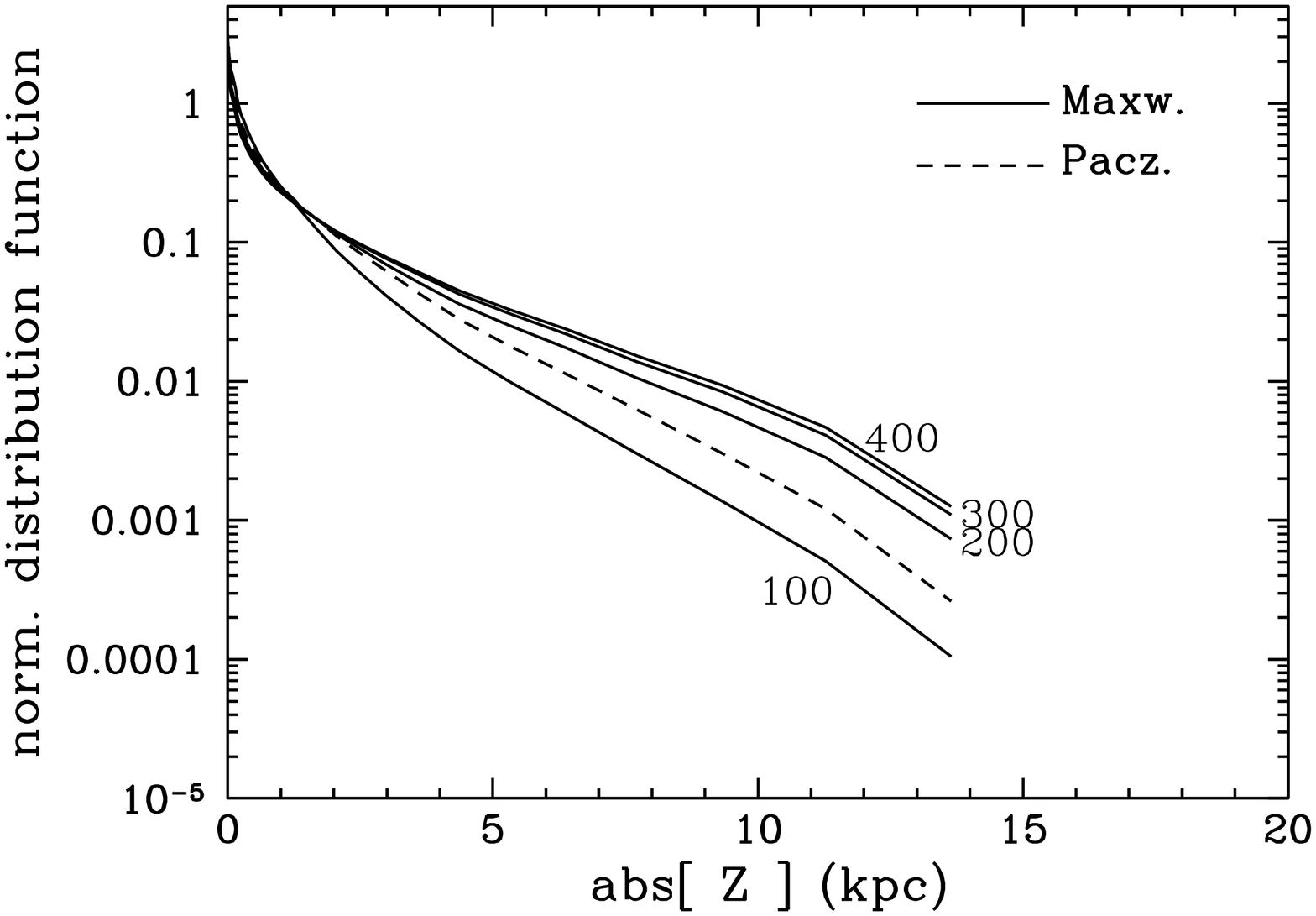,angle=-90,width=6.5in}}
 \figcaption{Probability distributions $f(\vert Z\vert)dZ$ of absolute
vertical distance $Z$ from the Galactic plane of coalescing NS--NS
binaries at present. A constant star--formation rate has been assumed for
the Galaxy. Line coding is as in Figure 1.}

 \centerline{\psfig{figure=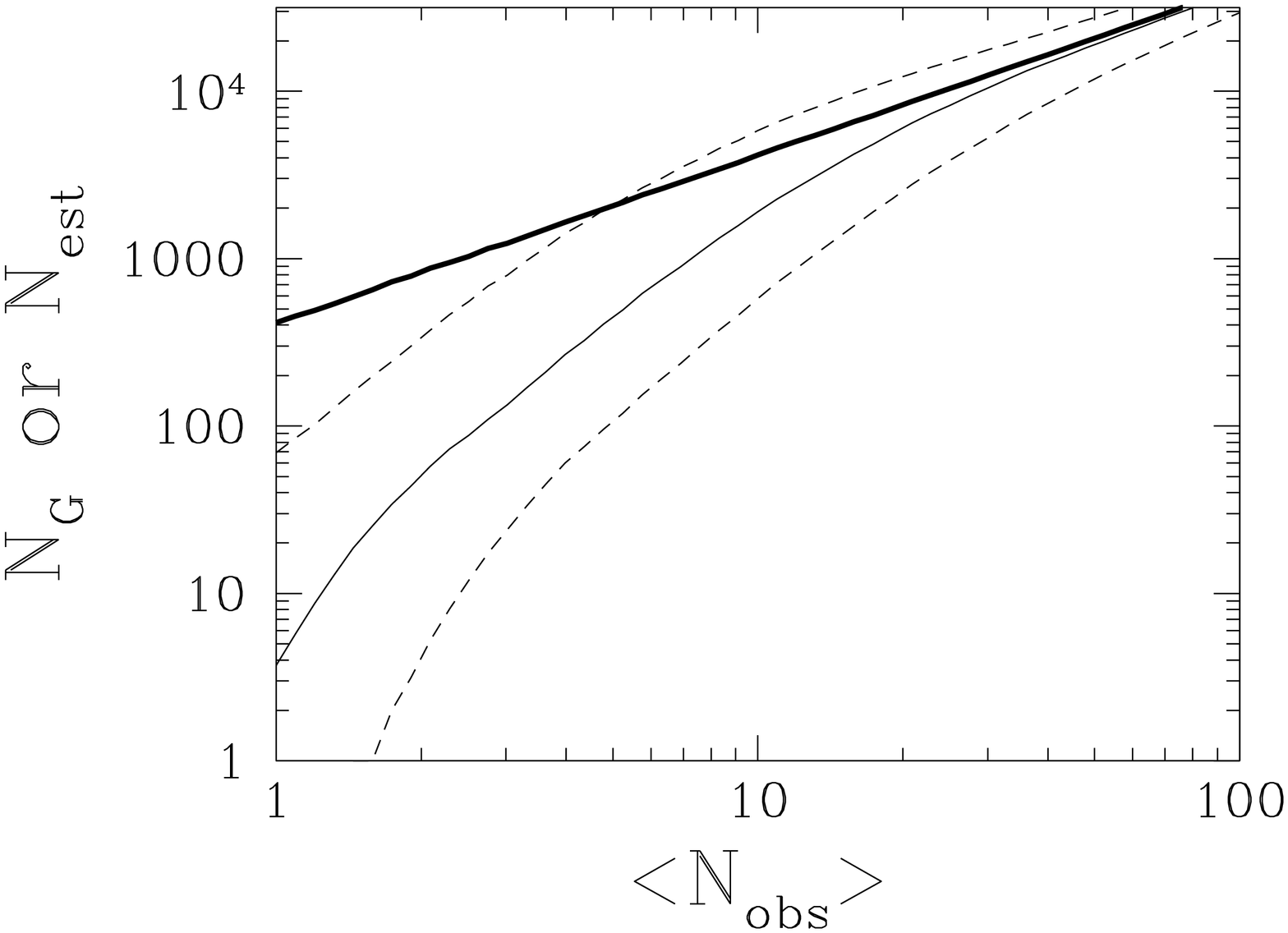,angle=-90,width=6.5in}}
 \figcaption{Total number $N_G$ of pulsars in the Galaxy (thick solid
line) and statistical properties of the distribution of the estimated
pulsar number $N_{\rm est}$ (median: thin, solid line; first quartile:
lower thin dashed line; third quartile: upper thin dashed line) as a
function of the mean number of objects in the observed samples. Note that
the mean estimated number $<N_{\rm est}>$ is equal to the total number
$N_G$ in the model. Curves are shown for our standard case: $p=2$, $L_{\rm
min}=1$\,mJy\,kpc$^2$, $L_1=30$\,mJy\,kpc$^2$, and
$L_2=3000$\,mJy\,kpc$^2$ (see text), which corresponds to Model \#1 in
Table 1.}

\centerline{\psfig{figure=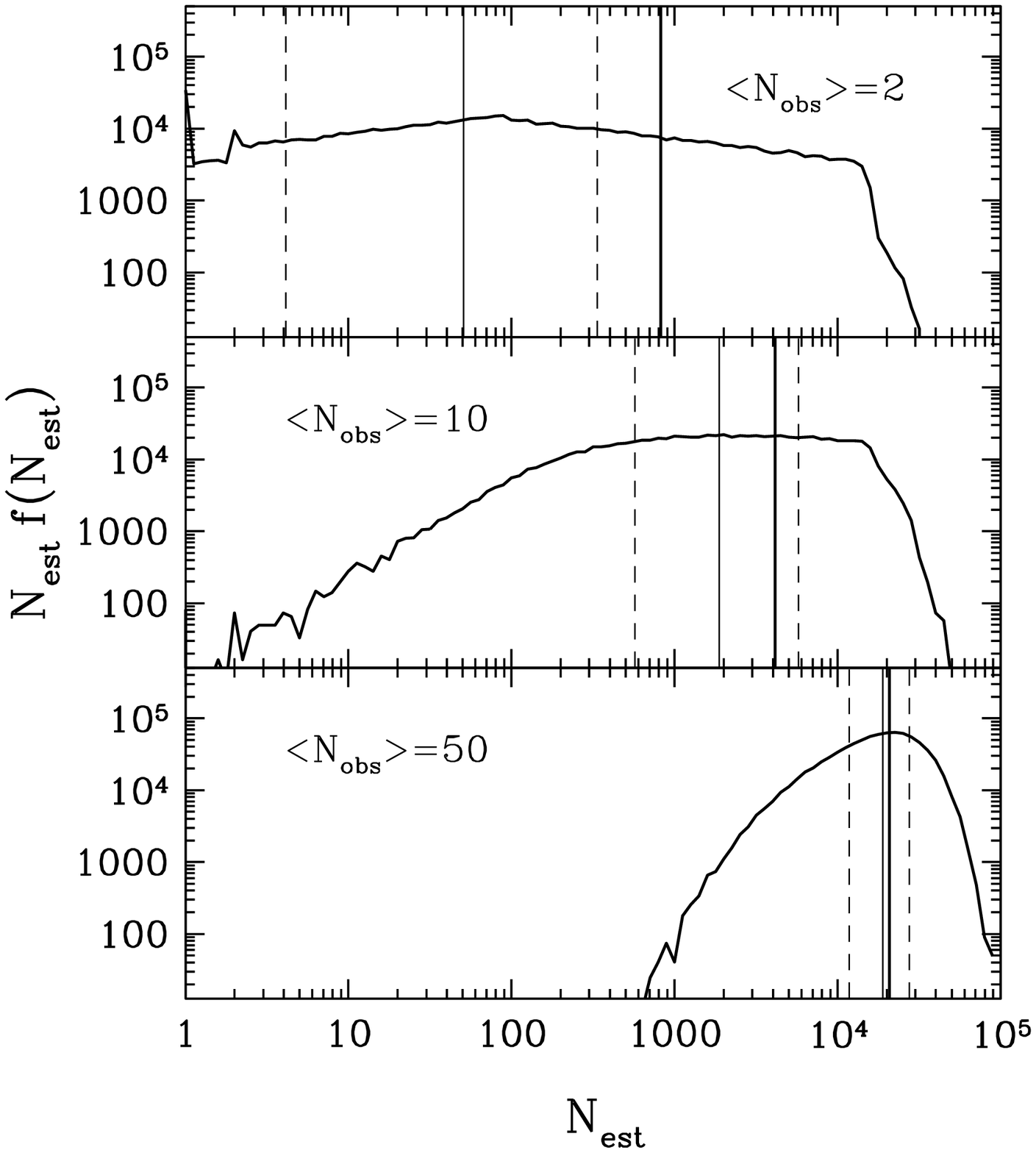,width=5.in}}
 \figcaption{Distribution functions $f(N_{\rm est})dN_{\rm est}$ of the
estimated number $N_{\rm est}$ of pulsars for $<N_{\rm obs}>=2,10,50$
(top, middle, and bottom panels, respectively). Vertical lines mark the
mean (thick solid line), median (thin solid line), and first and third
quartiles (left and right dashed lines, respectively) values of the
distributions. Results are shown for model number 1 (see Table 1).}

\centerline{\psfig{figure=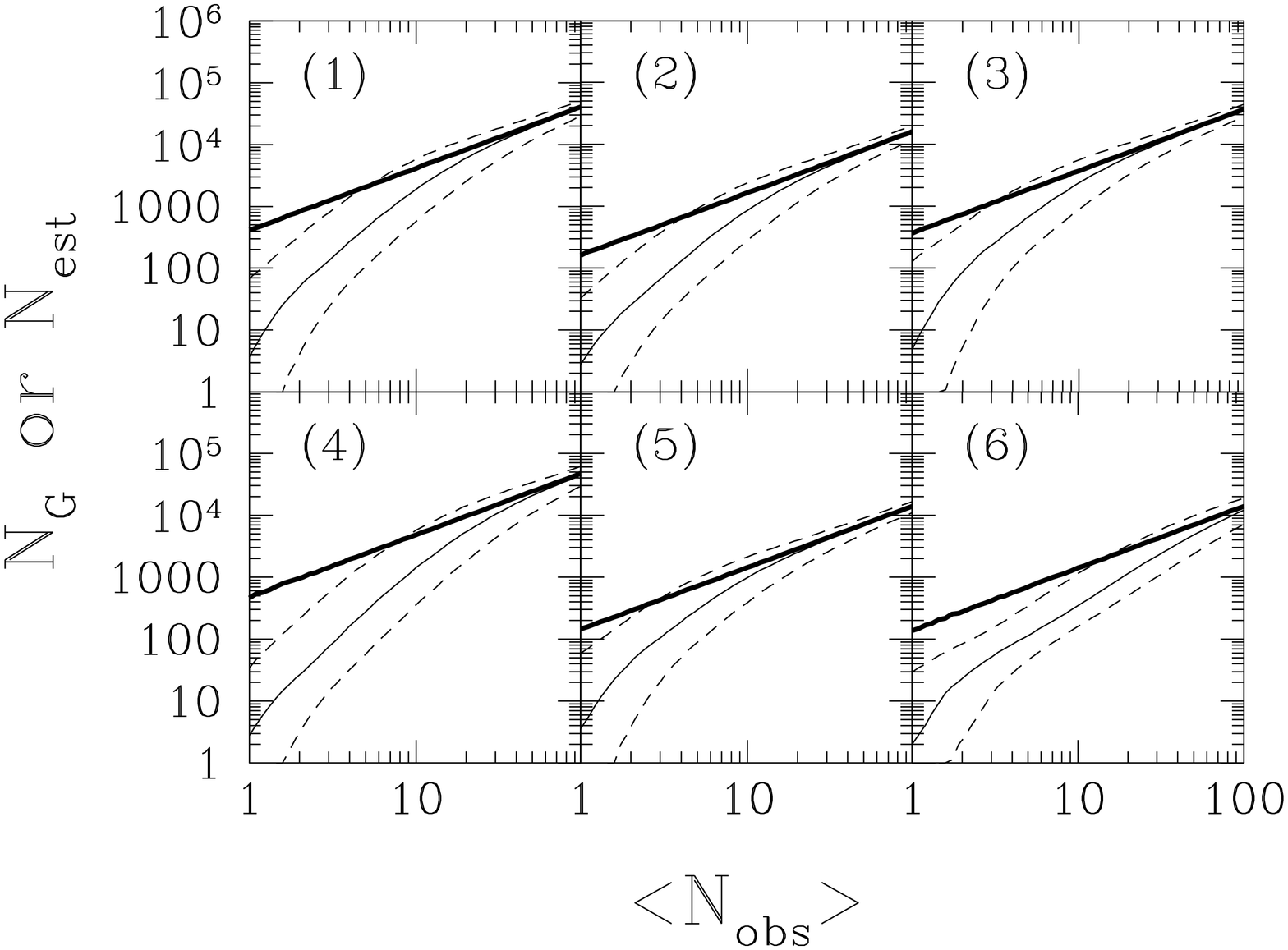,angle=-90,width=6.5in}}
\figcaption{Same as in Figure 4. Each panel shows curves for one of the
models listed in Table 1. Models are shown according to their model number
(see Table 1)  starting at the top left corner and ending at the bottom
right corner.}

\centerline{\psfig{figure=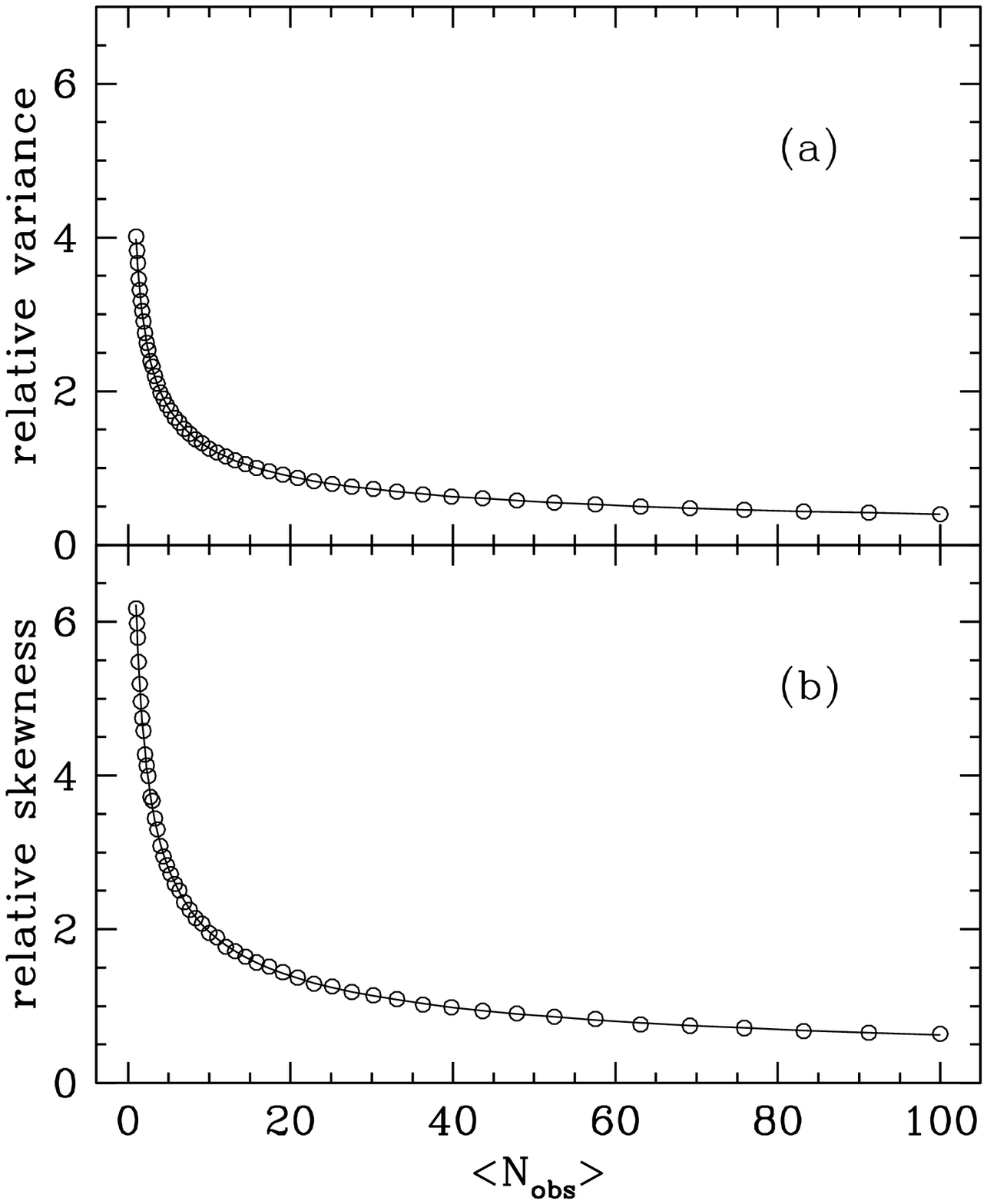,width=5.in}}
 \figcaption{Moments of the $N_{\rm est}$--distribution: (a) the variance 
relative to the mean and (b) the skewness relative to the variance as a
function of the mean number of observed objects, for
our standard case (model \#\,1 in Table 1). The points are the results of 
our Monte Carlo simulations and the solid curve is the analytical results 
(equations [A16] and [A17]). The agreement is better than 1\%. } 

\begin{deluxetable}{ccccccccccccc}
\tablewidth{500pt}
\tablecaption{Correction Factors based on Distribution Properties
of the Estimated Total Number of Pulsars $N_{\rm est}$ based on
{\em Model} Observed Samples}
\tablehead{ \multicolumn{5}{c}{Model Parameters} &
\multicolumn{4}{c}{$<N_{\rm obs}>=2$\tablenotemark{a}} &
\multicolumn{4}{c}{$<N_{\rm obs}>=3$\tablenotemark{a}} \\
\colhead{Number} & \colhead{p\tablenotemark{b}}
& \colhead{$L_{\rm min}$\tablenotemark{c}} &
\colhead{$L_1$\tablenotemark{c}} &
\colhead{$L_2$\tablenotemark{c}} & \colhead{$N_G$\tablenotemark{d}} &
\colhead{$\frac{N_G}{{\rm Median\tablenotemark{e}}}$} &
\colhead{$\frac{N_G}{{\rm 25\%\tablenotemark{f}}}$} &
\colhead{$\frac{N_G}{{\rm 75\%\tablenotemark{g}}}$} &
\colhead{$N_G$\tablenotemark{d}}
& \colhead{$\frac{N_G}{{\rm Median\tablenotemark{e}}}$} &
\colhead{$\frac{N_G}{{\rm 25\%\tablenotemark{f}}}$} &
\colhead{$\frac{N_G}{{\rm 75\%\tablenotemark{g}}}$}}

\startdata

1 & 2 & 1 & 30 & 3000 & 830 & 16 & 200 & 2.5 & 1245 & 9 & 52 & 1.5
\\
2 & 2 & 1 & 30 & 1000 & 325 & 13 & 107 & 2 & 490 & 7.5 & 35 & 1.3
\\
3 & 2 & 1 & 10 & 3000 & 745 & 8.5 & 143 & 1.5 & 1115 & 5 & 31 & 1
\\
4 & 2 & 1 & 100 & 3000 & 970 & 36 & 305 & 4.5 & 1450 & 20 & 102 & 2.5
\\
5 & 2 & 3 & 30 & 3000 & 290 & 7 & 75 & 1.3 & 430 & 4 & 21 & 0.95
\\
6 & 1.8 & 1 & 30 & 3000 & 270 & 12 & 120 & 3 & 400 & 8 & 32 & 2.8

\enddata

\tablenotetext{a}{Mean number of objects in the model observed sample}
\tablenotetext{b}{Power--law index of the pulsar luminosity function
($\propto L^{-p}$)}
\tablenotetext{c}{Break points in scale factor dependence on
luminosity in units of mJy\,kpc$^2$ (see eq.\ [16])}
\tablenotetext{d}{Model total number of pulsars in the Galaxy}
\tablenotetext{e}{Median of the $N_{\rm est}$ distribution}
\tablenotetext{f}{First quartile  of the $N_{\rm est}$ distribution}   
\tablenotetext{g}{Third quartile  of the $N_{\rm est}$ distribution}

\end{deluxetable}

\end{document}